\documentclass[twocolumn]{aastex7}
\usepackage{graphicx}
\usepackage{multirow}
\usepackage{rotating}
\usepackage{supertabular}
\usepackage{longtable}
\usepackage{CJK}
\usepackage[T1]{fontenc}
\usepackage[utf8]{inputenc}
\usepackage{tikz}
\usepackage{threeparttable}
\usepackage{hyperref}
\usepackage{makecell}
\usepackage{amsmath}
\usepackage{xcolor}
\usepackage{ulem}

\newcommand\gaia{Gaia}

\begin{document}
\begin{CJK*}{UTF8}{gbsn}
\title{A Comprehensive All-Sky Catalog of 3345 Molecular Clouds from Three-dimensional Dust Extinction}
\correspondingauthor{Haibo Yuan (苑海波)}
\email{yuanhb@bnu.edu.cn}

\author[0000-0002-4878-1227]{Tao Wang (王涛)}\email{wt@mail.bnu.edu.cn}
\affiliation{Institute for Frontiers in Astronomy and Astrophysics,
Beijing Normal University, Beijing, 102206, China; }
\affiliation{School of Physics and Astronomy, Beijing Normal University No.19, Xinjiekouwai St, Haidian District, Beijing, 100875, China; }

\author[0000-0003-2471-2363]{Haibo Yuan (苑海波)}\email{yuanhb@bnu.edu.cn}
\affiliation{Institute for Frontiers in Astronomy and Astrophysics,
Beijing Normal University, Beijing, 102206, China; }
\affiliation{School of Physics and Astronomy, Beijing Normal University No.19, Xinjiekouwai St, Haidian District, Beijing, 100875, China; }

\author[0000-0003-2472-4903]{Bingqiu Chen (陈丙秋)}\email{bchen@ynu.edu.cn}
\affil{South-Western Institute for Astronomy Research, Yunnan University, Kunming 650500, China;} 

\author[0000-0003-3144-1952]{Guangxing Li (李广兴)}\email{gxli@ynu.edu.cn}
\affil{South-Western Institute for Astronomy Research, Yunnan University, Kunming 650500, China;} 

\author[0000-0002-1259-0517]{Bowen Huang (黄博闻)}\email{huangbw@mail.bnu.edu.cn}
\affiliation{Institute for Frontiers in Astronomy and Astrophysics,
Beijing Normal University, Beijing, 102206, China; }
\affiliation{School of Physics and Astronomy, Beijing Normal University No.19, Xinjiekouwai St, Haidian District, Beijing, 100875, China; }

\author[0000-0001-5737-6445]{Helong Guo (郭贺龙)}\email{helong guo@mail.ynu.edu.cn}
\affil{South-Western Institute for Astronomy Research, Yunnan University, Kunming 650500, China;} 

\author[0000-0003-1863-1268]{Ruoyi Zhang (张若羿)}\email{zry@mail.bnu.edu.cn}
\affiliation{Institute for Frontiers in Astronomy and Astrophysics,
Beijing Normal University, Beijing, 102206, China; }
\affiliation{School of Physics and Astronomy, Beijing Normal University No.19, Xinjiekouwai St, Haidian District, Beijing, 100875, China; }

\begin{abstract}

Understanding the distribution and properties of molecular clouds is crucial for tracing the structure and evolution of the interstellar medium and the large-scale morphology of the Milky Way. Here we present an all-sky catalog of 3,345 molecular clouds identified from our previous three-dimensional dust reddening map using a dendrogram-based clustering method with distance-adaptive parameters. The catalog spans heliocentric distances from 90\,pc to 4.3\,kpc and includes key physical properties for each cloud, including position, size, mass, surface density, and dust density. Approximately 650 clouds in our catalog are associated with the boundary of the Local Bubble, while around 740 clouds (excluding those associated with the Local Bubble) are located at high Galactic latitudes ($|b| > 20^\circ$). The spatial distribution of the cataloged clouds reveals prominent large-scale features in the Galactic disk, including coherent spur-like structures, large-scale cavities, and a more detailed view of the Local Bubble shell. These findings refine our understanding of how molecular clouds trace the Galactic spiral arm network and provide new insight into the spatial structure of the Local Bubble. The catalog serves as a valuable resource for future studies of star formation, Galactic structure, and the interaction between molecular clouds and large-scale ISM features.

\end{abstract}

\section{Introduction} \label{sec:intro}

Despite accounting for only a small fraction of the Galaxy’s total mass, molecular gas plays a crucial role in Galactic evolution. Most of this gas is concentrated in molecular clouds, which are the primary sites of star formation \citep{1986ApJ...301..398M,1989ApJ...339..149S}. Understanding how molecular clouds form, evolve, and are distributed across the Galaxy is essential for studying both the origin of stars and the processes that shape Galactic structure and dynamics \citep{2013A&A...559A..34L,2016AA...591A...5L,2014ApJ...797...53G,2015MNRAS.450.4043W}.

Early identification of molecular clouds relied primarily in survey photographs, where molecular clouds appeared as dark patches obscuring background starlight \citep{1919ApJ....49....1B,1927cdosB}. Since the first detection of carbon monoxide (CO) emission by \cite{1970ApJ...161L..43W}, CO emission lines have become the main observational tool for mapping molecular gas distributions.

Large-scale CO surveys have significantly enhanced the ability to identify molecular clouds across the Galaxy. For instance, \citet{1985ApJ...295..402M} used observations from the 5-m Millimeter Wave Observatory to identify 57 clouds at high Galactic latitudes. Later, \citet{1997A&A...327..325M} used data from the 1.2-m Southern Millimeter-wave Telescope to identify 177 clouds in the southern outer Galaxy. \citet{2001ApJ...547..792D} compiled extensive CO datasets into comprehensive maps, which enabled systematic studies of molecular cloud properties by \citet{2016ApJ...822...52R} and \citet{Miville2017}. These studies applied algorithms such as dendrogram-based decomposition and Gaussian clustering to identify over 1000 and 8000 molecular clouds, respectively, across the Galactic disk.

While CO surveys have been instrumental, accurately determining molecular cloud distances remains a key challenge. Most distance estimates rely on kinematic models based on Galactic rotation curves \citep[e.g.,][]{1985ApJ...295..402M,1997A&A...327..325M,2016ApJ...822...52R,Miville2017} or associations with objects of known distances \citep[e.g.,][]{2008AN....329...10M}. These methods generally suffer from large uncertainties when applied to clouds that lack massive young stars or do not follow regular Galactic rotation. Additionally, deviations from circular motions and the near-far distance ambiguity further reduce the accuracy of kinematic distance estimates.

In addition, observations have revealed a population of ``CO-dark'' molecular clouds, which emit little or no CO but are detectable through molecular hydrogen and dust extinction \citep[e.g.,][]{2005Sci...307.1292G,2010ApJ...716.1191W,2011MNRAS.412..337G,2014MNRAS.441.1628S,2025NatAs}. This finding highlights the need for complementary observational techniques to build a more complete picture of molecular cloud populations.

Dust-based methods offer an alternative approach to study molecular clouds. Dense dust regions can be traced through stellar extinction, and when combined with accurate stellar parallax measurements \citep{2016AA...595A...1G}, this allows for reliable distance estimates. \citet{2019ApJ...879..125Z} adopted the extinction breakpoint method to estimate distances to local molecular clouds using \textit{Gaia} DR2 data. \citet{chen2020} used hierarchical clustering on three-dimensional (3D) reddening maps from \citet{chen2019} to identify 567 clouds within $|b| \leq 10^\circ$. Similarly, \citet{guo2022} applied the same methodology to identify 250 molecular clouds in the southern sky using extinction maps from \citet{guo2021}. \citet{Cahlon2024} applied a dendrogram-based method to the high-resolution 3D dust map of \citet{Leike2020}, identifying 65 molecular clouds within 116 -- 440 pc.. \citet{sun2024} identified 315 high-latitude clouds from Planck 857 GHz observations \citep{2020A&A...641A...1P}, with distances derived from stellar parallax and extinction. More recently, \citet{xie2024} used the 3D dust map of \citet{Vergely2022} and identified 550 molecular clouds within 3 kpc of the solar neighborhood.


Building on previous studies of the distribution and properties of molecular clouds in the Milky Way, we construct a new all-sky catalog of 3345 molecular clouds identified using the deep, high-resolution, all-sky 3D dust extinction map from \citeauthor{WangT2025} (2025, ApJS, in press; hereafter W25). The catalog spans heliocentric distances from 90 pc to 4.3 kpc and includes key physical properties for each cloud, such as position, size, mass, surface density, and dust density. Approximately 650 clouds in our catalog are associated with the boundary of the Local Bubble—a low-density cavity in the interstellar medium thought to have formed through the action of multiple supernovae, and surrounded by a shell of swept-up dust and gas \citep{1987ARA&A..25..303C,2003A&A...411..447L,2009Ap&SS.323....1W,2022Natur.601..334Z}. Around 740 additional clouds, excluding those related to the Local Bubble, are located at high Galactic latitudes ($|b| > 20^\circ$). Each cloud entry is accompanied by distance estimates and key physical parameters, offering a valuable resource for future studies of star formation and Galactic structure. The structure of the paper is as follows: Section 2 introduces the dataset, Section 3 details our method for identifying molecular clouds and determining their physical parameters from the three-dimensional extinction map, Section 4 presents our results, Section 5 discusses these findings, and Section 6 provides a summary.

\section{Data} \label{sec:data}

This study uses the 3D dust reddening map published by W25, which provides an all-sky view of the reddening distribution. The map was constructed by combining reddening measurements from two major sources: low-resolution spectra from the Large Sky Area Multi-Object Fiber Spectroscopy Telescope (LAMOST; \citealt{2012RAA....12.1197C, 2012RAA....12..723Z, 2014IAUS..298..310L}) and BP/RP (hereafter XP) spectra from the European Space Agency's (\textit{ESA}) \textit{Gaia} mission \citep{2016AA...595A...1G}. 

Specifically, the $E(B-V)_{\rm LAMOST}$ values were derived using the standard star-pair technique \citep{2013MNRAS.430.2188Y} applied to approximately 4.6 million unique stars. These estimates were based on stellar parameters from LAMOST DR11 and synthetic $B$/$V$-band photometry generated from \textit{Gaia} XP spectra, resulting in a typical precision of about 0.01 mag.

The $E(B-V)_{\rm XP}$ values, taken from the catalog by \citet{zhang2023}, were obtained through forward modeling of \textit{Gaia} XP spectra. These measurements were cross-matched with $E(B-V)_{\rm LAMOST}$ to validate their accuracy, resulting in a high-quality sample of around 150 million stars. The combined dataset achieves a median precision of approximately 0.03 mag in $E(B-V)$.

To construct the 3D map, W25 modeled the relationship between reddening and distance along different lines of sight using smooth, monotonically increasing parametric functions. This parametric model accounts for contributions from the Local Bubble, the diffuse interstellar medium, and various molecular cloud components. The sky was divided adaptively based on local stellar density, yielding angular resolutions that range from 3.4$^{\prime}$ to 58$^{\prime}$. About half of the sky achieves a resolution finer than 6.9$^{\prime}$. 

The map's reddening estimates for individual stars have a typical precision of about 0.01 mag in regions with Galactic latitudes $|b| > 20^\circ$. In lower-latitude areas ($|b| < 20^\circ$), the precision ranges between 0.01 and 0.05 mag. The map reaches maximum distances of 3--5 kpc in high-extinction regions near the Galactic plane ($|b| < 5^\circ$), and extends out to 10--15 kpc in less obscured regions.

In this work, we use the Python package \texttt{dustmaps3d}\footnote{The \texttt{dustmaps3d} package, which provides programmatic access to the W25 3D dust map, is available via \texttt{pip install dustmaps3d}. Its source code and documentation are hosted on GitHub at \url{https://github.com/Grapeknight/dustmaps3d}, and an archived version is available at the NADC (\url{https://nadc.china-vo.org/res/r101619}). The package has been assigned a persistent identifier: \dataset[DOI: 10.12149/101619]{https://doi.org/10.12149/101619}.
} to access the W25 3D dust map, apply a Gaussian smoothing with a $\sigma$ of 0.1$^\circ$ on the plane of the sky, and subsequently identify molecular cloud structures through a hierarchical clustering approach. The 3D dust map used in this study is publicly available from the website: \url{https://nadc.china-vo.org/data/dustmaps}.

\section{Method} \label{sec:method}

Molecular gas in the interstellar medium displays a clear hierarchical structure. High-density regions are typically small and embedded within larger, lower-density environments \citep[e.g.,][]{1986ApJ...300L..89B,1992ApJ...393L..25L}. The diffuse gas occupies most of a molecular cloud’s volume and forms the base of this hierarchy, while dense cores at the top serve as the primary sites for star formation \citep{1998ApJ...504..223G,2004A&A...416..191T,2008ApJ...672..410L}.

To capture and analyze these nested structures, hierarchical clustering provides a natural and effective framework. Dendrograms, which represent hierarchical relationships between structures, are especially useful for interpreting the internal complexity of molecular clouds. In this study, we use the open-source \texttt{astrodendro} package in \texttt{Python} \citep{Rosolowsky2008} to extract and analyze these structures from three-dimensional dust maps.

\subsection{Constructing 3D Data Cubes from All-Sky Data}\label{subsec:Cubes}

To apply \texttt{Dendrograms} to the W25 3D dust extinction map, we first convert the all-sky data into a set of processable 3D data cubes. Directly gridding the map in Galactic longitude-latitude coordinates would introduce projection distortions, especially near the Galactic poles. The W25 map employs the HEALPix scheme \citep{HEALPix}, which divides the sky into equal-area pixels. The resolution is controlled by the parameter $\rm N_{side} = 2^k$, where $k$ is an integer. This results in $12 \times \rm N_{side}^2$ pixels covering the full sky. Although W25 uses adaptive pixel sizes, all pixels can be uniformly subdivided to $\rm N_{side} = 1024$ ($k = 10$), which we adopt in this work.

Since HEALPix pixels are not aligned with Galactic meridians, we adopt a remapping strategy to ensure spatial consistency (see Figure~\ref{fig:HEALPix}). The sky is divided into 12 base pixels at $k = 0$, and each is refined independently to $k = 10$, producing $1024 \times 1024$ grids.  Each high-resolution HEALPix pixel is mapped to 2D coordinates $(X^{\prime}, Y^{\prime})$ using the nested ordering scheme:

\begin{figure*}
    \centering
    \includegraphics[width=1\linewidth]{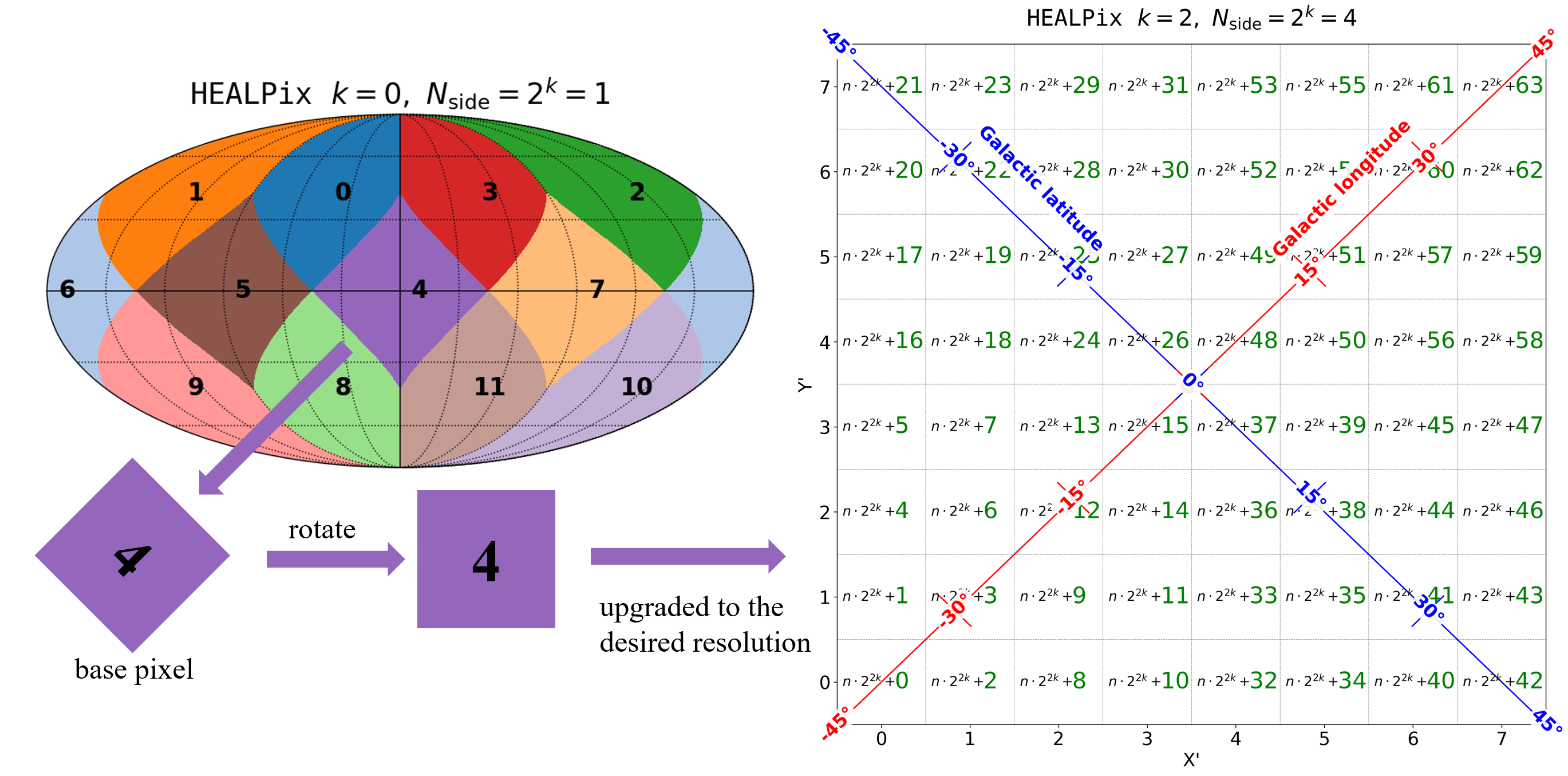}
    \caption{Left: the 12 base HEALPix pixels at $k = 0$ resolution dividing the full sky into equal-area regions. Right: an example of refining the $n = 4$ base pixel to $k = 2$, resulting in an $8 \times 8$ grid. The main and secondary diagonals correspond to Galactic longitude and latitude of $0^\circ$, spanning $90^\circ$. The numbers in each cell represent the HEALPix pixel IDs (with nested ordering).}
    \label{fig:HEALPix}
\end{figure*}

\begin{equation}
\mathrm{pixid_{nested}} - n \times 2^{2k}  = \sum_{i=0}^{m} d_i \times 4^{i}, \quad d_i \in \{0,1,2,3\}, 
\label{eq:1}
\end{equation}

where $n$ is the base pixel index ($0$–$11$), $k = 10$ is the resolution level, and $m = \left\lfloor \log_{4}{(\mathrm{pixid_{nested}} - n \times 2^{2k})}  \right\rfloor$ is the highest digit in the base-4 expansion. The pixel’s 2D coordinates are then given by:

\begin{equation}
X^{\prime} = \sum_{i=0}^{m} \left\lfloor \frac{d_i}{2} \right\rfloor \times 2^{i},
\label{eq:2}
\end{equation}
\begin{equation}
Y^{\prime} = \sum_{i=0}^{m} (d_i \mod 2) \times 2^{i}.
\label{eq:3}
\end{equation}

This transformation enables distortion-free reconstruction of the data cube and prepares the input for \texttt{Dendrograms}-based clustering.

\begin{figure}
    \centering
    \includegraphics[width=1\linewidth]{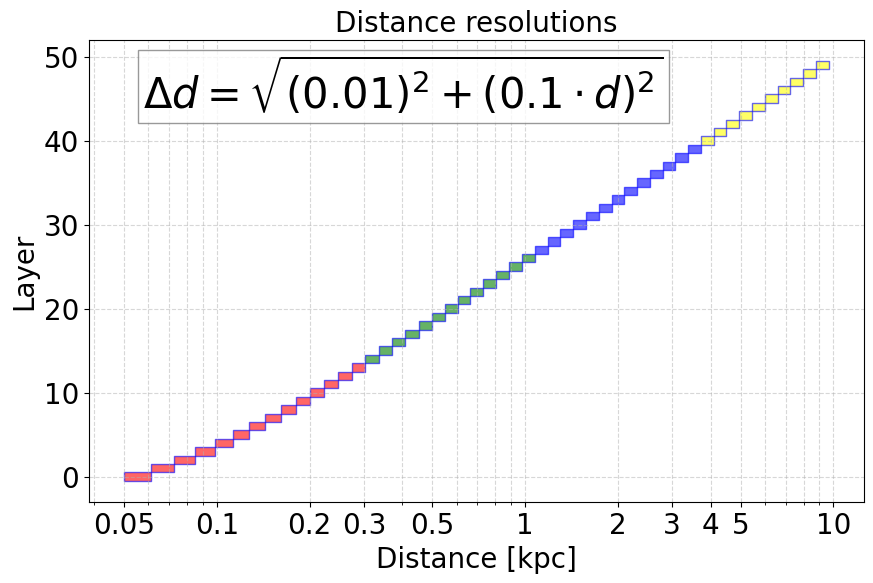}
    \caption{Bin widths as a function of distance. Colors indicate the use of different parameters in different regions, as discussed in Section~\ref{subsec:DENDROGRAM}.}
    \label{fig:distance_resolutions}
\end{figure}

We discretize the third dimension—distance—using an adaptive binning scheme to account for increasing uncertainties at larger distances (see Figure~\ref{fig:distance_resolutions}):

\begin{equation}
\Delta d = \sqrt{(0.01)^2 + (0.1 \cdot d)^2}.
\end{equation}

We begin binning at 0.05 kpc to exclude the Local Bubble and continue up to 10 kpc, resulting in 50 non-uniform distance bins. For each bin, we compute the extinction gradient:

\begin{equation}
\Delta E(B-V) = \frac{E(B-V)_{d_\mathrm{up}} - E(B-V)_{d_\mathrm{low}}}{d_\mathrm{up} - d_\mathrm{low}},
\end{equation}

where $d_\mathrm{low}$ and $d_\mathrm{up} = d_\mathrm{low} + \Delta d$ are the lower and upper bounds of the bin. This extinction gradient (in units of mag/kpc) is assigned as the voxel value in the three-dimensional data cube.

We construct 12 data cubes (one per base HEALPix pixel), each with 50 distance layers and $1024 \times 1024$ spatial resolution. These serve as input for the hierarchical clustering analysis using \texttt{Dendrograms}.

\subsection{Identification of the Molecular clouds}\label{subsec:DENDROGRAM}

\texttt{Dendrograms} identify hierarchical structures using iso-contours of intensity. The algorithm requires three parameters: (1) \texttt{min\_value}, the minimum intensity threshold; (2) \texttt{min\_delta}, the minimum contrast between a substructure and its parent; and (3) \texttt{min\_npix}, the minimum number of connected pixels for a structure. Because distant clouds tend to be smaller and fainter due to projection effects and uncertainties, we apply different parameter sets to specific distance intervals (see Table~\ref{tab:param}). 

These parameter choices were guided by extensive empirical testing across multiple trials.\footnote{Figure~\ref{fig:param1234} in the supplementary material presents histograms of cloud physical properties derived using several different clustering parameter sets.} The \texttt{min\_value} parameter is primarily used to exclude noise below a specified threshold. We tested values in the range \texttt{min\_value} = 0.01 -- 0.05 mag/kpc and found that the clustering results are largely insensitive to this choice. However, lower thresholds significantly increase computational costs. Therefore, we adopt a uniform value of \texttt{min\_value} = 0.05 mag/kpc across all distance bins. The \texttt{min\_delta} parameter controls the minimum contrast required between a substructure and its parent structure. If set too low, particularly in the 0.3 -- 1 kpc distance bin, the algorithm may misidentify noise fluctuations at high Galactic latitudes as cloud structures. Conversely, in more distant bins, setting \texttt{min\_delta} too high can suppress the identification of real structures, as extinction contrast tends to be lower. To balance these effects, we adopt distance-dependent values of 0.5, 0.4, 0.3, and 0.1 mag/kpc for progressively increasing distance bins. The \texttt{min\_npix} parameter defines the minimum number of connected pixels required for a structure to be considered valid. The values we adopt correspond to angular diameters of approximately $3^\circ$, $1.5^\circ$, $1^\circ$, and $0.5^\circ$, which translate to physical sizes of roughly 10, 20, 30, and 40~pc, respectively. These scales are comparable to those of typical molecular clouds.

Although each parameter set is run on the full cube, only structures whose centroids fall within the corresponding distance range are retained for analysis.

\setlength{\tabcolsep}{1.5mm}{
\begin{table*}[htbp]
\footnotesize
\centering
\caption{Parameter configurations for \texttt{Dendrograms} clustering tailored to different distance intervals. Each configuration is applied to the full data cube, but only the structures within the indicated distance range are selected for analysis.}
\begin{tabular}{ccccccc}
\hline
\hline
Distance Range & Number of Distance Bins & min\_value & min\_delta & min\_npix & \makecell{Angular Diameter \\ (from min\_npix)} & \makecell{Physical Diameter \\ (from min\_npix)} \\
\hline
\hline
$<$ 0.3 kpc       & 14 & 0.05 mag/kpc & 0.5 mag/kpc & 2000 & $\sim$3.0 deg & $\sim$10 pc \\
0.3 -- 1 kpc      & 13 & 0.05 mag/kpc & 0.4 mag/kpc & 500  & $\sim$1.5 deg & $\sim$20 pc\\
1 -- 3.5 kpc      & 13 & 0.05 mag/kpc & 0.3 mag/kpc & 200  & $\sim$1.0 deg & $\sim$30 pc\\
$>$ 3.5 kpc       & 10 & 0.05 mag/kpc & 0.1 mag/kpc & 50   & $\sim$0.5 deg & $\sim$40 pc\\
\hline
\hline
\end{tabular}
\label{tab:param}
\end{table*}}

We focus on identifying the \textit{leaves} of the dendrograms, which are the smallest, non-divisible structures corresponding to high-density regions. These are adopted as molecular clouds in our analysis. While dendrograms also represent larger, nested structures called branches, our study concentrates on the most compact components.\footnote{See the \texttt{Dendrograms} documentation, Section `Plotting contours of structures in third-party packages', particularly Figures 1 and 2: \url{https://dendrograms.readthedocs.io/en/stable/plotting.html}} We did not apply special treatment to clouds located at the boundaries between the 12 3D cubes. This is because the \texttt{Dendrograms} method naturally decomposes large molecular cloud structures into multiple substructures (e.g., dense cores). For clouds that are split at cube boundaries, each fragment typically contains only a portion of the original cloud’s mass and size, and thus this splitting does not significantly affect the estimation of parameters such as surface density. Moreover, such cases are relatively rare, accounting for only about 5\% of the total sample, with approximately 70\% of them located within 0.3 kpc.

\subsection{Determination of Sky Coordinates, Distances, and Physical Parameters}\label{sec:parameters}

\begin{figure}
    \centering
    \includegraphics[width=1\linewidth]{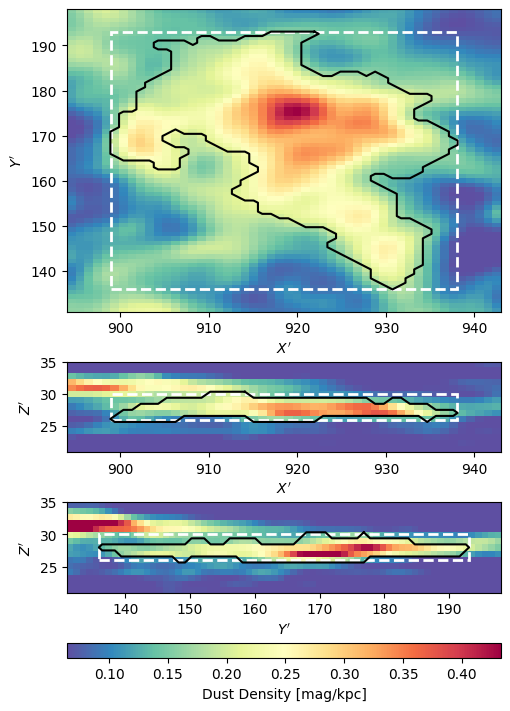}
    \caption{Example projections of a dendrogram clustering result. Top: $X^{\prime}$--$Y^{\prime}$ plane (sky view), Middle: $X^{\prime}$--$Z^{\prime}$ plane, Bottom: $Y^{\prime}$--$Z^{\prime}$ plane. Here, $Z^{\prime}$ represents the line-of-sight (distance) direction, and $X^{\prime}$--$Y^{\prime}$ corresponds to the celestial sphere, as described in Section~\ref{subsec:Cubes}. The color scale represents the average dust density (i.e., extinction gradient in mag/kpc). For each projection, the mean is calculated using pixels enclosed by the white dashed rectangle, which are selected to minimize contamination from overlapping lines of sight.}
    \label{fig:dendrogram-example}
\end{figure}

For each \texttt{Dendrograms} leaf (see Figure~\ref{fig:dendrogram-example}), we extract all voxel indices $(X^{\prime}, Y^{\prime}, Z^{\prime})$ and convert them back to Galactic coordinates $(l, b, d)$ using the inverse of Equations~\ref{eq:1} - \ref{eq:3}. Each voxel holds a dust density value, defined as the local extinction gradient in mag\,kpc$^{-1}$.

We compute a dust-weighted centroid for each cloud as:

\begin{equation}
\bar{q} = \frac{\sum_i q_i \cdot \mathrm{dust}_i}{\sum_i \mathrm{dust}_i},
\end{equation}

where $q_i$ is the position in $l$, $b$, or $d$, and $\mathrm{dust}_i$ is the extinction gradient. The centroid $(\bar{l}, \bar{b}, \bar{d})$ is used to represent the spatial position of each molecular cloud in the subsequent analysis. To accurately compute $\bar{l}$, the periodic nature of Galactic longitude is taken into account by averaging the values in radians and then converting the result to degrees within the range $[0^\circ, 360^\circ)$.

Given that our analysis is based on the HEALPix pixelization scheme, each pixel on the sky covers an identical solid angle. The solid angle per pixel is calculated as:

\begin{equation}
\Omega_{\mathrm{pix}} = \frac{4\pi}{12 \times (2^k)^2},
\end{equation}

with $k = 10$. For each molecular cloud, we first count the number of unique pixels on the sky ($N_{\mathrm{pix}}$) associated with the dendrogram leaf. The total solid angle of the cloud is then:

\begin{equation}
\Omega = N_{\mathrm{pix}} \cdot \Omega_{\mathrm{pix}}.
\end{equation}

Using the cloud's centroid distance $\bar{d}$, we compute its projected area and physical radius:

\begin{equation}
S = \Omega \bar{d}^2, \quad r = \sqrt{\frac{S}{\pi}},
\end{equation}

To estimate mass, we convert each voxel’s extinction gradient to color excess:

\begin{equation}
E(B-V)_i = \mathrm{dust}_i \cdot \Delta d_i.
\end{equation}

Following \citet{2011A&A...535A..16L,2015ApJ...799..116S} and \citet{2017MNRAS.472.3924C}, the total mass is:

\begin{equation}
M = \frac{\mu m_{\mathrm{H}}}{\mathrm{DGR}} \sum_{i=1}^{N} \Omega_{\mathrm{pix}} \cdot d_i^2 \cdot A_{V, i},
\end{equation}

where $\mu = 1.37$ \citep{2011A&A...535A..16L} is the mean molecular weight per hydrogen atom (assuming a standard interstellar medium composition of 63\% hydrogen, 36\% helium, and 1\% dust), $m_{\mathrm{H}} = 1.674 \times 10^{-24}\,\mathrm{g}$ is the mass of a hydrogen atom, and $\mathrm{DGR}$ is the dust-to-gas ratio:

\begin{equation}
\mathrm{DGR} = \frac{A_{V}}{N(\mathrm{H})}=\frac{A_{V}}{N\left(\mathrm{H}_{\mathrm{I}}\right)+2 N\left(\mathrm{H}_{2}\right)} .
\end{equation}

adopting $\mathrm{DGR} = 4.15 \times 10^{-22}~\mathrm{mag\,cm^2}$ \citep{2017MNRAS.472.3924C}. The corresponding $V$-band extinction $A_{V, i}$ is then obtained assuming a standard total-to-selective extinction ratio $R_V = 3.1$.

The surface mass density $\Sigma$ is:

\begin{equation}
\Sigma = \frac{M}{S}.
\end{equation}

The average dust density is $\bar{\rho}_{\mathrm{dust}}$:

\begin{equation}
\bar{\rho}_{\mathrm{dust}}= \frac{1}{N} \sum_{i=1}^{N} \mathrm{dust}_i,
\end{equation}

where $\mathrm{dust}_i$ is the extinction gradient (in mag/kpc) in the $i$-th pixel.

\section{Results} \label{sec:results}

\begin{figure*}
    \centering
    \includegraphics[width=1\linewidth]{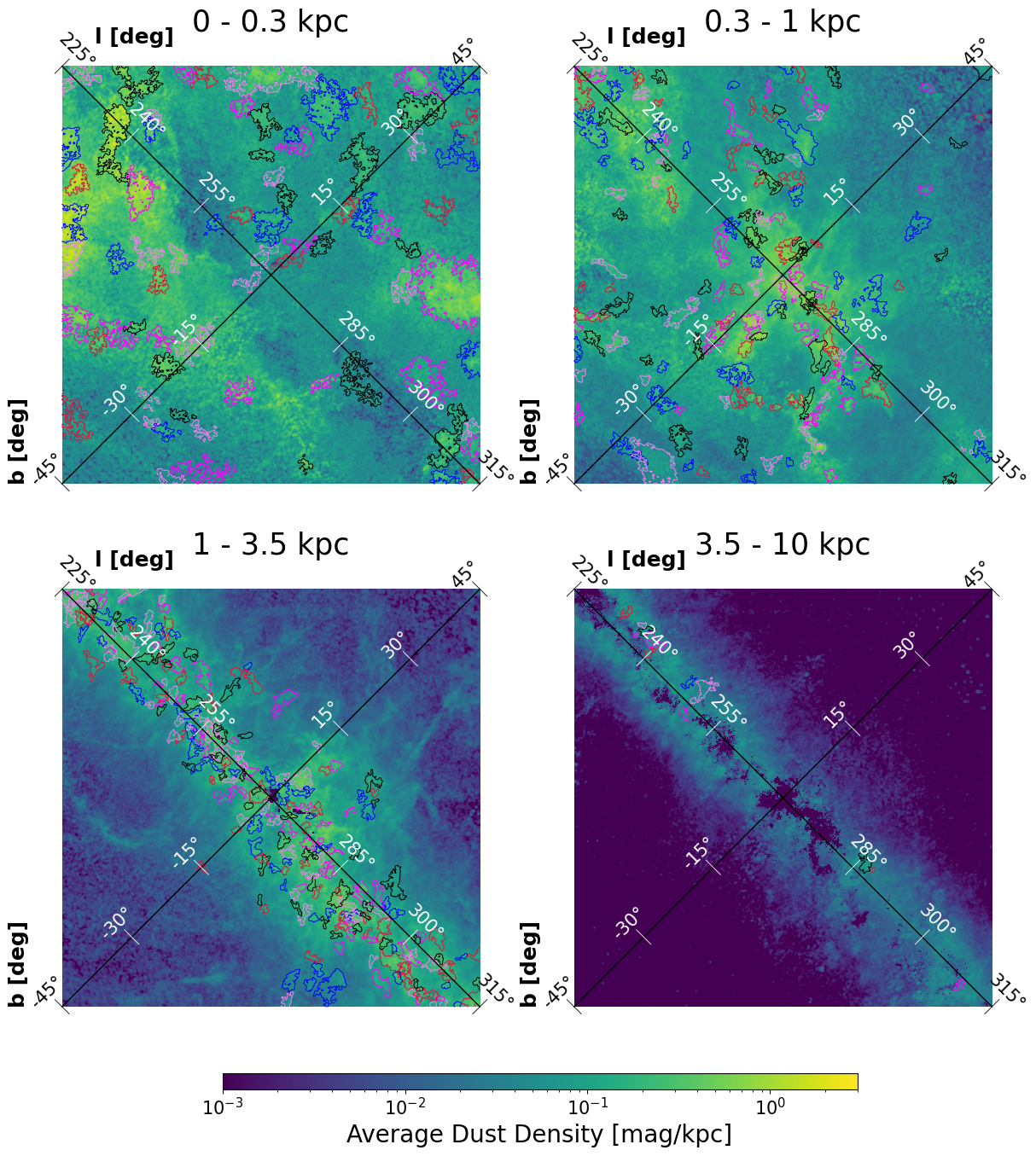}
    \caption{
    This figure presents the clustering results for the region corresponding to HEALPix pixel index 7 in Figure \ref{fig:HEALPix}, spanning Galactic longitudes from $225^\circ$ to $315^\circ$ and latitudes from $-45^\circ$ to $+45^\circ$. The four subpanels correspond to different distance intervals: 0--0.3\,kpc, 0.3--1\,kpc, 1--3.5\,kpc, and 3.5--10\,kpc. Molecular clouds identified in each bin are outlined accordingly. The \texttt{Dendrograms} parameters used in each distance interval are listed in Table~\ref{tab:param}. Distinct clouds along the line of sight are represented using different colors.
    }
    \label{fig:Results-example1}
\end{figure*}

\begin{figure*}
    \centering
    \includegraphics[width=1\linewidth]{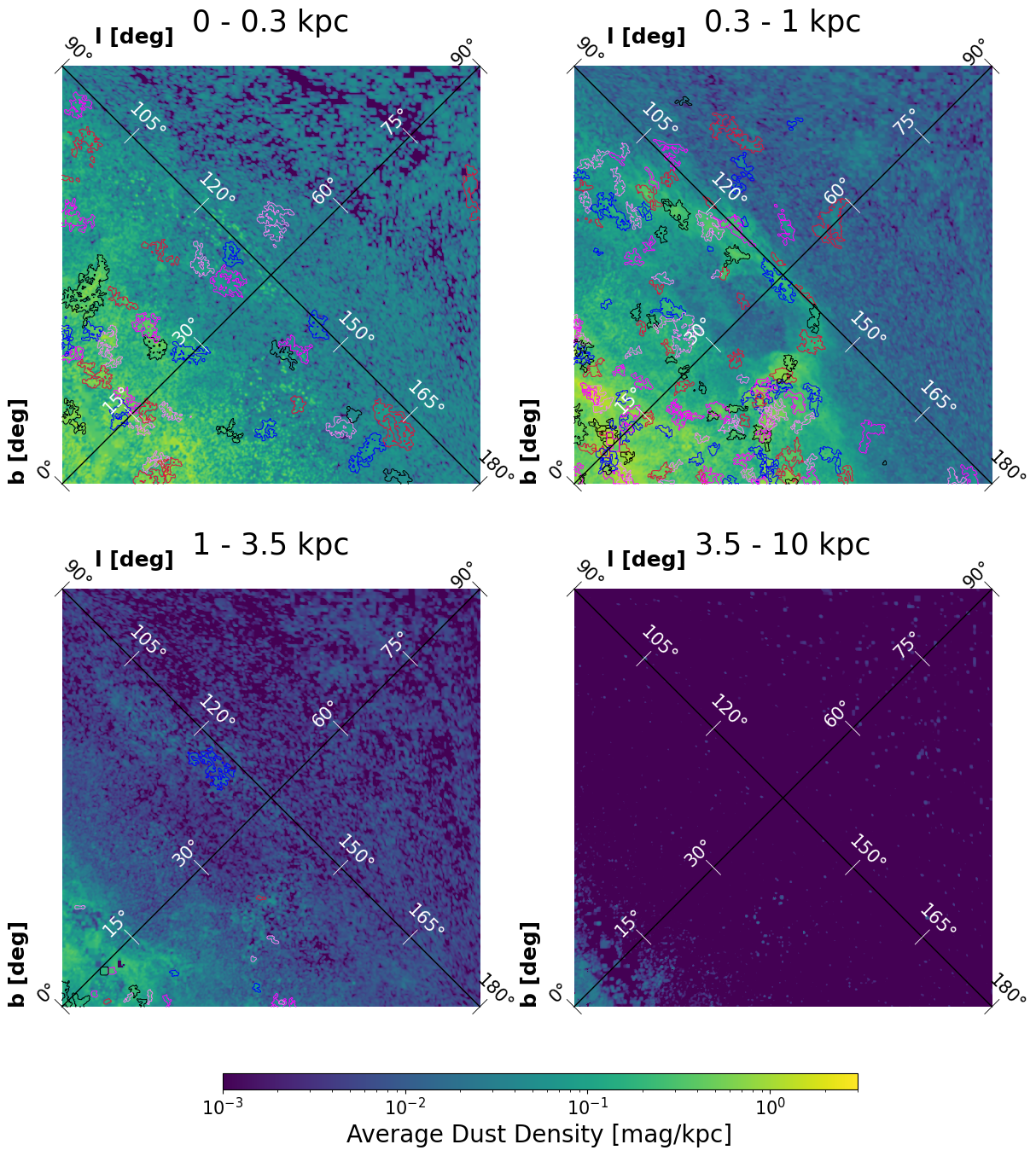}
    \caption{Same as Figure~\ref{fig:Results-example1}, but for the region spanning Galactic longitudes $90^\circ$ to $180^\circ$ and latitudes $0^\circ$ to $90^\circ$, corresponding to HEALPix pixel index 1 in Figure~\ref{fig:Results-example1}.
    }
    \label{fig:Results-example2}
\end{figure*}

Using the \texttt{Dendrograms} algorithm applied to the W25 3D dust reddening map, we identified a total of 3345 molecular clouds across the entire sky. Examples of clustering results for two representative sky regions are shown in Figures~\ref{fig:Results-example1} and \ref{fig:Results-example2}. The clustering outcomes for the remaining ten sky sectors—completing the twelve-part division adopted in this analysis—are provided in the supplementary material.

The full catalog is presented in Table~\ref{catalogue}, where each row lists the main properties of an individual molecular cloud. These include Galactic coordinates ($l$, $b$), distance ($d$), solid angle ($\Omega$), physical radius ($r$), total mass ($M$), surface mass density ($\Sigma$), average dust density ($\bar{\rho}_{\mathrm{dust}}$), and a flag indicating whether the cloud lies along the boundary of the Local Bubble.

\begin{table*}
  \centering
  \caption{Catalog of molecular clouds.}
  \begin{tabular}{rrrr  *{6}{r} }
    \hline
    \hline
    ID  & $l$        & $b$       & $d$       & $r$  & $\Omega$          & $M$           & $\Sigma$               & $\bar{\rho}_{\mathrm{dust}}$ & Local Bubble Flag \\
        & ($\degr$)  & ($\degr$) & (kpc) & (pc) & (deg$^2$)             & ($M_{\odot}$) & ($M_{\odot}$\,pc$^{-2}$) & (mag/kpc) & 1 or 0\\
    \hline
    1 & $41.05$ & $-71.64$ & $0.09$ & $1.65$ & $3.53$ & $4.11 \times 10^{0}$ & $0.48$ & $0.21$ & 1 \\
    2 & $44.64$ & $-79.71$ & $0.09$ & $2.32$ & $6.57$ & $4.81 \times 10^{0}$ & $0.28$ & $0.16$ & 1 \\
    3 & $21.90$ & $4.08$ & $0.11$ & $3.12$ & $8.91$ & $3.98 \times 10^{2}$ & $13.00$ & $7.78$ & 1 \\
    4 & $169.68$ & $-69.66$ & $0.11$ & $2.47$ & $5.54$ & $7.17 \times 10^{0}$ & $0.37$ & $0.16$ & 1 \\
    5 & $249.21$ & $50.92$ & $0.11$ & $3.84$ & $12.82$ & $1.99 \times 10^{1}$ & $0.43$ & $0.24$ & 1 \\
... & ... & ... & ... & ... & ... & ... & ... & ... & ... \\
    3341 & $292.47$ & $1.92$ & $4.08$ & $47.17$ & $1.38$ & $6.29 \times 10^{4}$ & $9.00$ & $0.16$ & 0 \\
    3342 & $117.79$ & $1.77$ & $4.18$ & $78.06$ & $3.60$ & $1.55 \times 10^{5}$ & $8.10$ & $0.12$ & 0 \\
    3343 & $300.52$ & $2.55$ & $4.19$ & $78.47$ & $3.61$ & $2.53 \times 10^{5}$ & $13.07$ & $0.26$ & 0 \\
    3344 & $108.98$ & $16.65$ & $4.28$ & $24.94$ & $0.35$ & $2.38 \times 10^{4}$ & $12.18$ & $0.28$ & 0 \\
    3345 & $188.11$ & $-0.30$ & $4.30$ & $24.61$ & $0.34$ & $9.81 \times 10^{3}$ & $5.15$ & $0.11$ & 0 \\
    \hline
  \end{tabular}
  \parbox{\textwidth}{\footnotesize \baselineskip 3.8mm
    {\it Note.} The Local Bubble flag indicates whether the cloud is located on the boundary of the Local Bubble (1 = yes, 0 = no).}
  \label{catalogue}
\end{table*}

\subsection{Spatial Distribution of Molecular Clouds}\label{sec:Spatial}

\begin{figure*}
    \centering
    \includegraphics[width=1\linewidth]{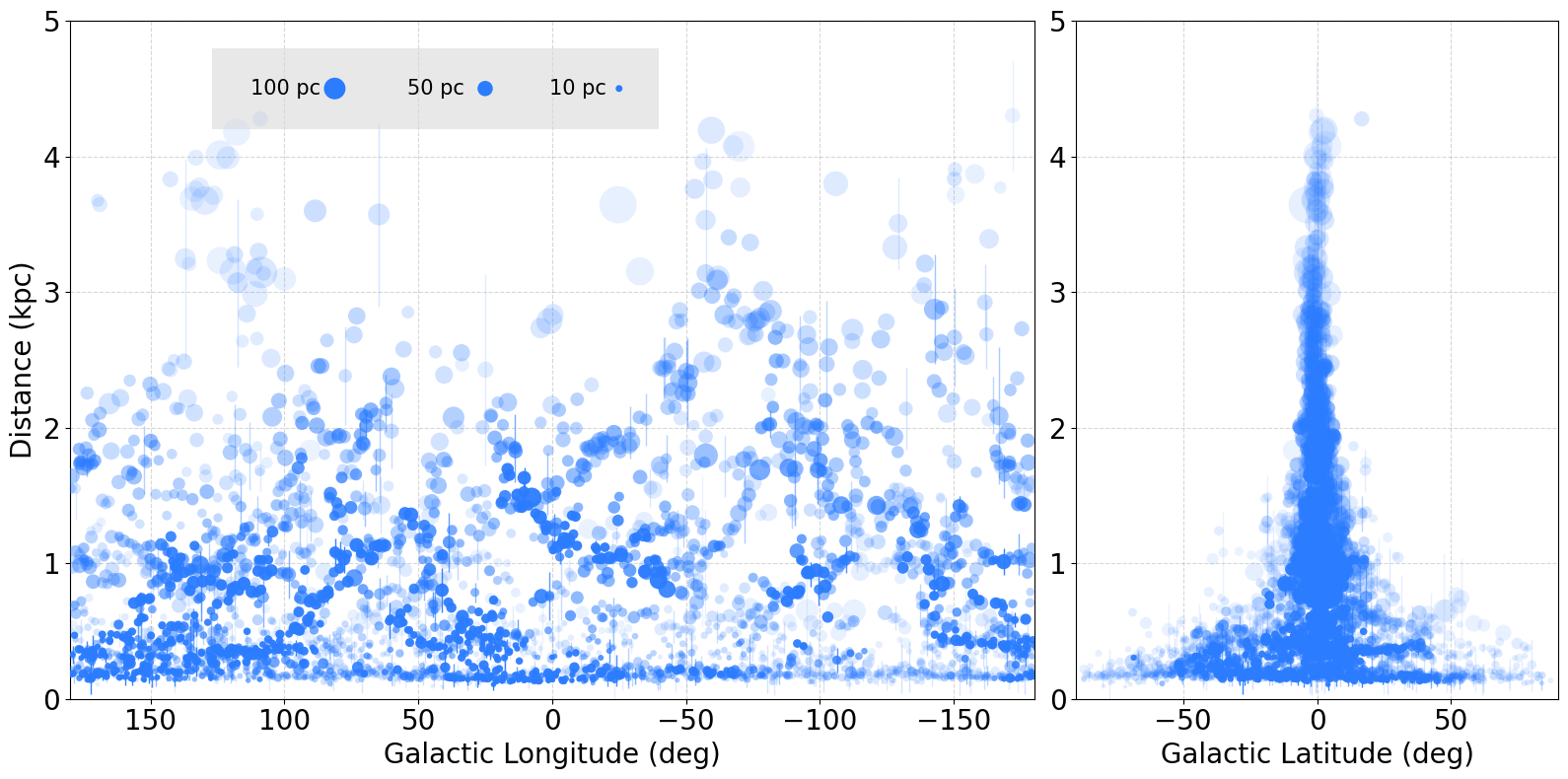}
    \caption{
    Distribution of molecular clouds in Galactic coordinates.
    Left: Galactic longitude versus distance from the Sun. Right: Galactic latitude versus distance. In both panels, each circle represents a molecular cloud. The radius of the circle reflects the physical size of the cloud, while its opacity indicates the average dust density—denser clouds appear more opaque.
    }
    \label{fig:MC-lb}
\end{figure*}

Figure~\ref{fig:MC-lb} presents the distribution of all identified molecular clouds in Galactic longitude ($l$), latitude ($b$), and distance ($d$). The clouds span a wide range of distances, from approximately 80\,pc to 4.5\,kpc. Their distribution follows the large-scale structure of the Galactic disk and is broadly consistent with the spiral arm features described in \citet{chen2020}.

\begin{figure*}
    \centering
    \includegraphics[width=1\linewidth]{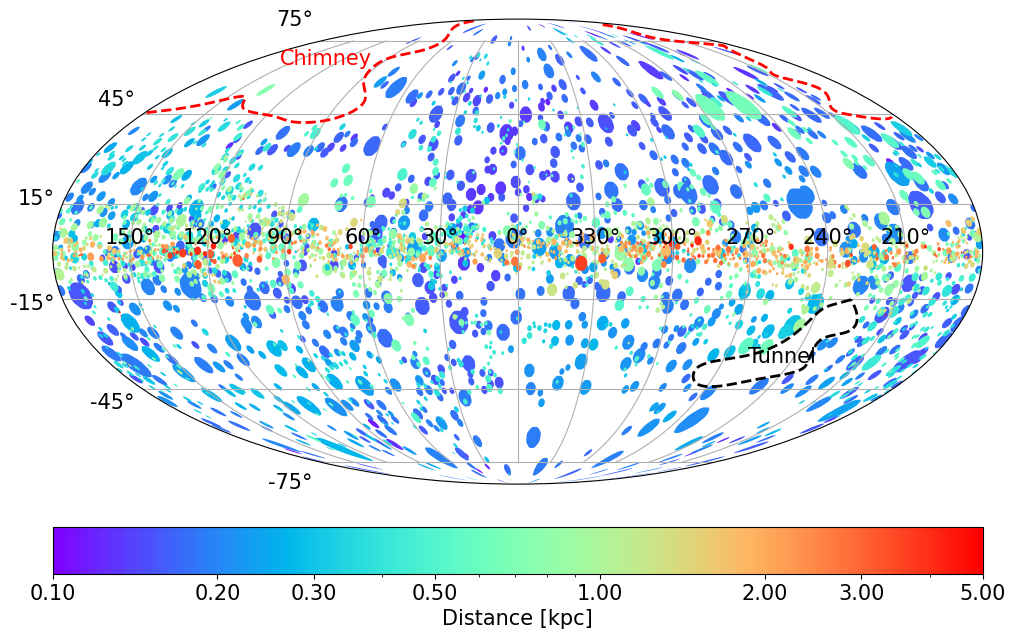}
    \caption{Sky distribution of cataloged molecular clouds in Galactic coordinates. Circle sizes represent angular radii, and colors indicate distances from the Sun. The void outlined by the red dashed line corresponds to the "chimney" structure identified by \citet{Neill2024}, interpreted as a vertical opening of the Local Bubble. The "tunnel" feature described in the same study is outlined by the black dashed line.
    }
    \label{fig:mollweid}
\end{figure*}

Figure~\ref{fig:mollweid} shows the sky distribution of clouds in Galactic coordinates using a Mollweide projection. The circle size reflects the angular radius derived from each cloud’s projected area, and the color indicates the distance from the Sun. Most high-latitude clouds ($|b| > 10^\circ$) are nearby, at distances less than 300\,pc. A clear void near Galactic longitude $l \sim 120^\circ$ toward the North Galactic Pole aligns with the "chimney" structure reported by \citet{Neill2024}. Both the chimney and tunnel features described in their study are apparent in our results, supporting the idea that the Local Bubble opens toward high latitudes. In the southern sky, we also identify several additional voids, which may represent other openings in the Local Bubble.

\begin{figure*}
    \centering
    \includegraphics[width=1\linewidth]{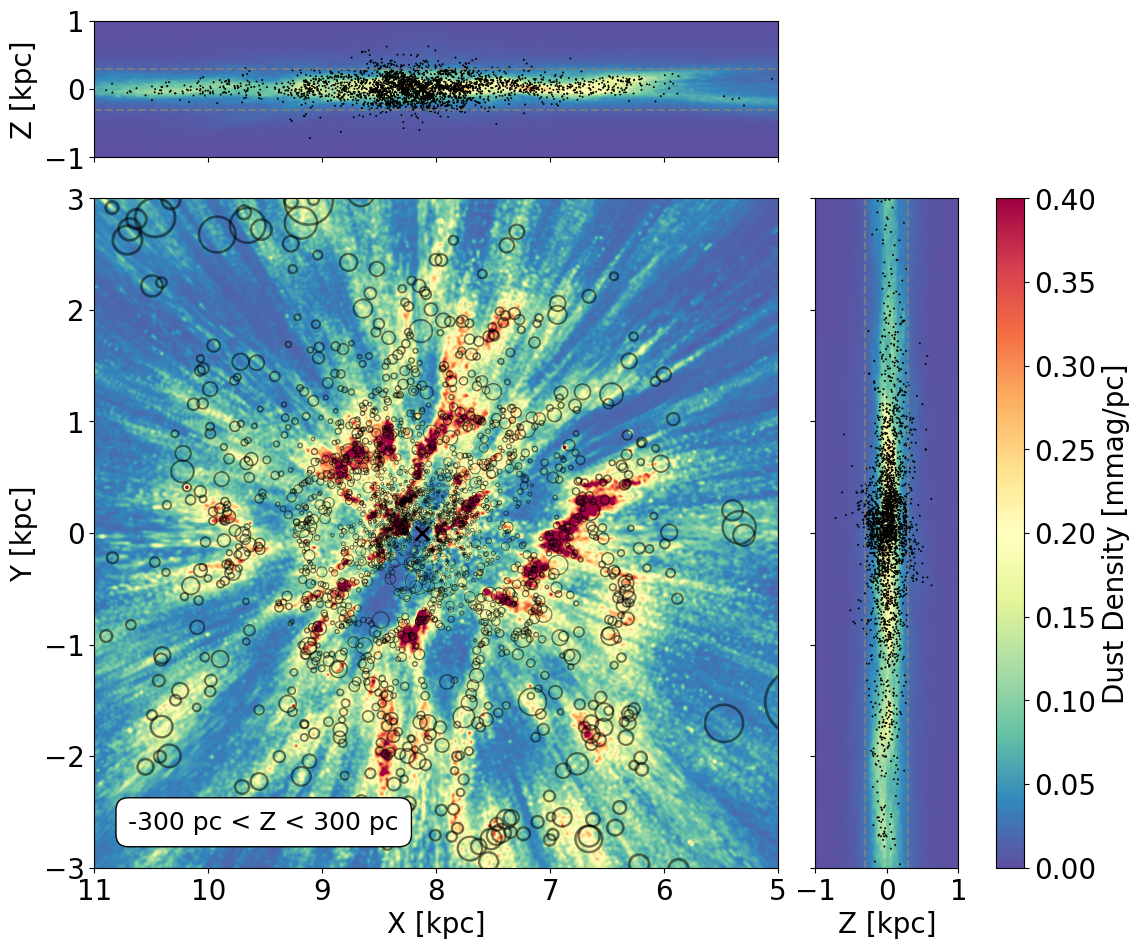}
    \caption{
    Spatial distribution of molecular clouds not associated with the Local Bubble in the $X$-$Y$, $X$-$Z$, and $Y$-$Z$ planes.The Sun is located at $(X, Y, Z) = (8.12, 0.00, 0.02)$\,kpc, and the Galactic center at the origin. In the $X$-$Y$ panel, clouds are shown as hollow circles with radii scaled to twice their actual physical radii to enhance visibility. Background shading reflects average dust density from the W25 map.The $X$-$Z$ and $Y$-$Z$ panels show clouds as black dots over the corresponding vertical dust distribution.
   }
    \label{fig:disk}
\end{figure*}

Figure~\ref{fig:disk} shows the spatial distribution of molecular clouds projected onto the $X$-$Y$, $X$-$Z$, and $Y$-$Z$ planes. Molecular clouds associated with the Local Bubble are excluded from this figure for clarity. The remaining clouds trace the high-density dust structures visible in the W25 map.

\begin{figure*}
    \centering
    \includegraphics[width=1\linewidth]{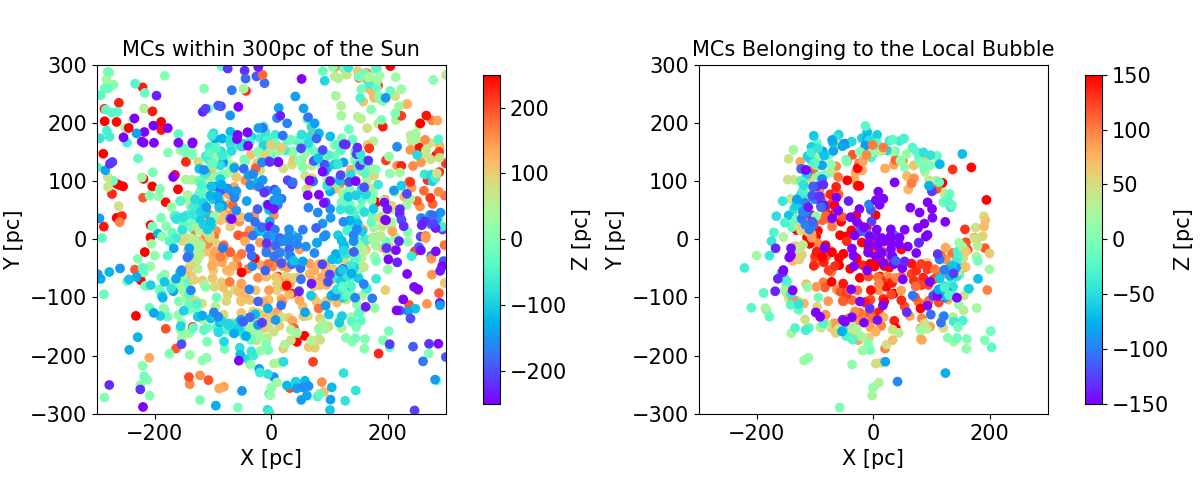}
    \caption{
    Left: spatial distribution of molecular clouds within 0.3\,kpc of the Sun, projected onto the $X$-$Y$ plane.  The Sun is at the origin.  The color scale shows the vertical coordinate $Z$ in parsecs.  Right: molecular clouds likely associated with the Local Bubble boundary, selected using the distance threshold described in the text.  Colors again represent $Z$.
    }
    \label{fig:Bubble}
\end{figure*}

Figure~\ref{fig:Bubble} displays the distribution of molecular clouds within 0.3\,kpc of the Sun, shown in Cartesian coordinates. In this coordinate system, the Sun is placed at the origin $(X, Y, Z) = (0, 0, 0)$. The left panel reveals a prominent quasi-spherical structure, which serves as an ideal tracer of the Local Bubble.

Figure~\ref{fig:Bubble} focuses on molecular clouds within 0.3\,kpc of the Sun, shown in Cartesian coordinates centered on the Sun. The left panel reveals a nearly spherical distribution of clouds, tracing the Local Bubble boundary. For each nearby cloud, we compute the average Local Bubble boundary distance based on the W25 extinction profile fit. Adding a conservative offset of 70\,pc, we classify clouds located within this limit as part of the Local Bubble shell. These selected clouds are shown in the right panel. A more detailed analysis of the Local Bubble and its surroundings will be presented in a separate work.

\subsection{Physical properties of the molecular clouds}\label{sec:Physical}

The molecular clouds identified in this study exhibit a wide range of physical properties. Cloud radii span approximately 2 to 100\,pc, with a median of about 8\,pc. Total masses range from $\sim$10\,$M_\odot$ to $6 \times 10^5$\,$M_\odot$, with a median near 1000\,$M_\odot$. Surface mass densities vary between 0.5 and 40\,$M_\odot$\,pc$^{-2}$, with a median of 6\,$M_\odot$\,pc$^{-2}$. Average dust densities range from roughly 0.1 to 10\,mag\,kpc$^{-1}$, with a median value of 0.7\,mag\,kpc$^{-1}$.

\begin{figure*}
    \centering
    \includegraphics[width=1\linewidth]{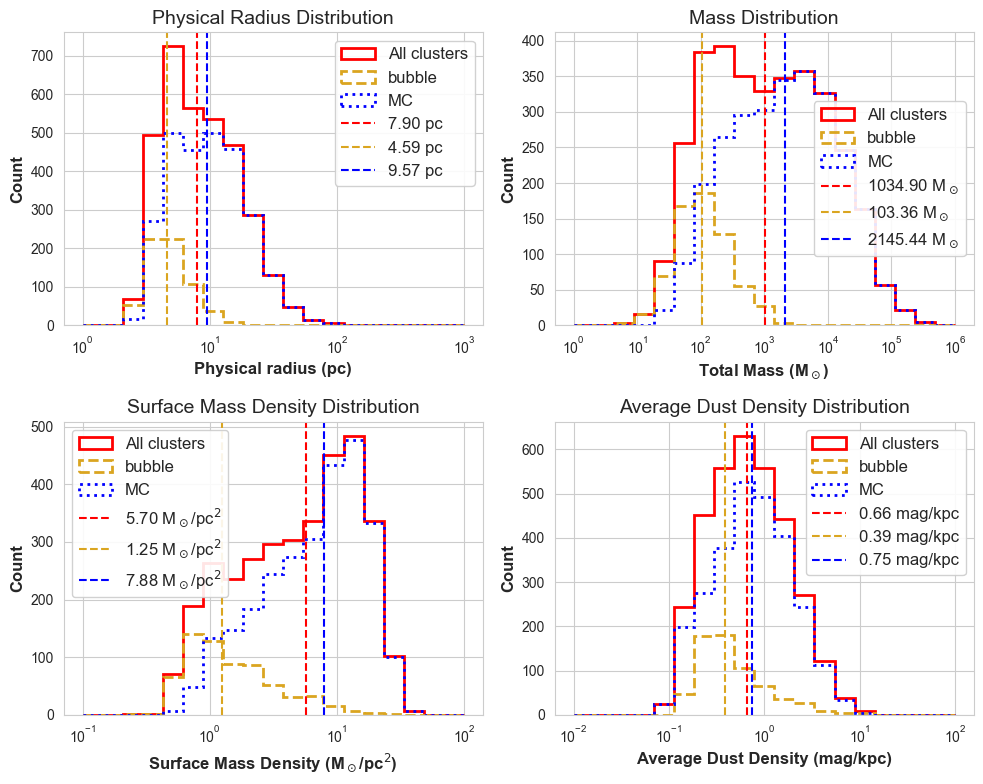}
    \caption{
    Histograms of key physical properties of the molecular clouds cataloged in this work. Top left: physical radius; top right: total mass; bottom left: surface mass density; bottom right: average dust density. Solid red lines represent the full sample, dashed goldenrod lines denote Local Bubble clouds, and dashed blue lines correspond to non-Local Bubble clouds.
    }
    \label{fig:this-work-histogram}
\end{figure*}

To explore differences between molecular clouds associated with the Local Bubble and those located elsewhere, we present histograms of key physical quantities in Figure~\ref{fig:this-work-histogram}. Systematic differences are evident between the two populations. For physical radius and total mass, variations are partly due to the use of distance-dependent \texttt{Dendrograms} parameters, which are tuned to optimize detection at different distances. In contrast, the observed differences in surface mass density likely reflect intrinsic physical differences. Many Local Bubble-associated clouds are less dense and possibly not in virial equilibrium, explaining their lower surface densities. Supporting this, \citet{2025NatAs} reported the discovery of a nearby dark molecular cloud named Eos, located just 94 pc from the Sun. Based on their analysis, Eos is expected to dissipate through photoevaporation within approximately 5.7 million years, indicating that it is not in virial equilibrium. In our sample, we identify Eos as corresponding to several clouds likely associated with the Local Bubble, visible in the top-left panel of Figure~\ref{fig:pix0}, spanning the region around Galactic coordinates (l $\sim$ 50$^{\circ}$, b $\sim$ 45$^{\circ}$). Cases like Eos suggest that some Local Bubble-associated clouds are transient structures shaped primarily by external pressure and radiative processes, rather than by self-gravity. Despite their low density, these clouds are valuable tracers of the Local Bubble’s structure.

\section{Discussion} \label{sec:discussion}

\subsection{Comparison of Physical Properties with Previous Studies}\label{sec:Comparison}

\begin{figure}
    \centering
    \includegraphics[width=1\linewidth]{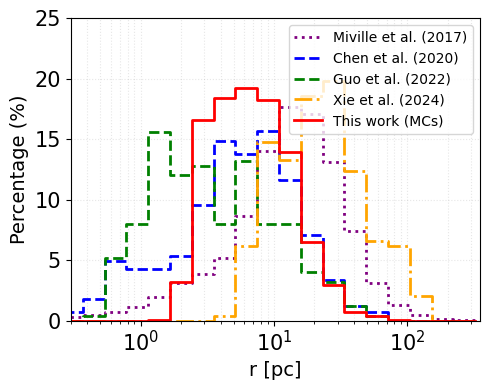}
    \includegraphics[width=1\linewidth]{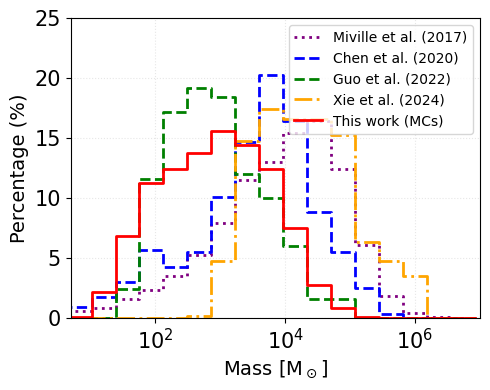}
    \includegraphics[width=1\linewidth]{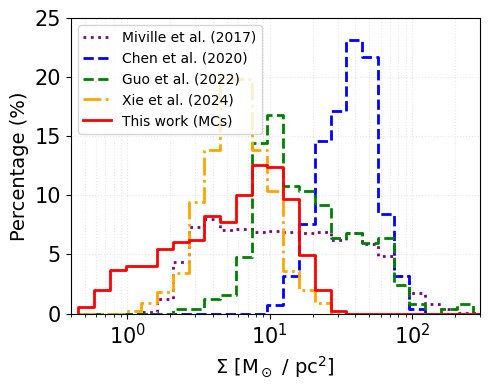}
    \caption{
    Comparison of physical properties of molecular clouds in this work (excluding Local Bubble clouds; red) with those from \citet{Miville2017} (purple), \citet{chen2020} (blue), \citet{guo2022} (green), and \citet{xie2024} (orange). Top: physical radius; middle: total mass; bottom: surface mass density. All plots show normalized histograms for comparison.
    }

    \label{fig:comparison}
\end{figure}

Figure~\ref{fig:comparison} compares the distributions of physical radius, mass, and surface mass density between clouds from this work (excluding Local Bubble clouds) and those identified by \citet{Miville2017}, \citet{chen2020}, \citet{guo2022}, and \citet{xie2024}. As these previous studies generally lack coverage of nearby clouds ($d < 0.2$\,kpc), we exclude Local Bubble structures for consistency.

\citet{Miville2017} identified 8107 molecular clouds using CO data from \citet{2001ApJ...547..792D}, relying on kinematic distances for physical parameter estimates. \citet{chen2020} and \citet{guo2022} used 3D dust maps and stellar distances to identify 567 and 250 clouds, respectively, using dendrogram-based clustering in $(l, b, d)$ space. \citet{xie2024} identified 550 molecular clouds within 3\,kpc using a 3D Cartesian dust map from \citet{Vergely2022}.

The cloud radii in our catalog are comparable to those in \citet{chen2020} and \citet{guo2022}, but systematically smaller than those in \citet{Miville2017} and \citet{xie2024}. This difference reflects the focus of our work on compact, dense regions, whereas the latter studies include more extended structures. The similarity to \citet{chen2020} and \citet{guo2022} is expected, as we also use the \texttt{Dendrograms} algorithm in $(l, b, d)$ space. However, our radius distribution is more concentrated, with fewer very small or very large clouds. This results from our use of distance-adaptive clustering parameters, which maintain consistent physical size selection across distances. In contrast, applying a single fixed set of parameters tends to select smaller clouds nearby and larger ones at greater distances.

Our mass estimates are systematically lower than those in \citet{Miville2017}, \citet{chen2020}, and \citet{xie2024}, though for different reasons. In \citet{Miville2017} and \citet{xie2024}, the higher masses stem from identifying large-scale cloud complexes. Our method, in contrast, targets denser substructures, which naturally leads to lower total masses.

Compared to \citet{chen2020} and \citet{guo2022}, our catalog yields roughly half the cloud masses. This is partially due to the limited number of massive disk clouds in the southern sky in \citet{guo2022}, but primarily stems from differences in mass estimation methods. In both \citet{chen2020} and \citet{guo2022}, dendrogram clustering was performed in $(l, b, d)$ space with fixed binning (0.2\,kpc), and total extinction was derived from fitting stellar extinction profiles along sightlines, including extended cloud material. In contrast, we sum extinction only within voxels identified in the dendrogram leaves, focusing on more dense regions. This approach results in lower total masses but yields a more physically meaningful measure of the gravitationally bound component. With the availability of high-resolution extinction maps such as W25, we recommend using voxel-based mass estimation methods that focus on dense, spatially resolved cloud structures, rather than relying on integrated extinction along the line of sight.

Our surface mass densities are lower than those in \citet{chen2020}, but similar to those in \citet{xie2024}. The lower values in our catalog can be attributed to two main factors. First, the reduction in total mass—due to voxel-based mass estimation—naturally leads to lower surface densities. Second, our sample includes a substantial number of less dense, high-latitude clouds ($|b| > 20^\circ$), even after excluding Local Bubble clouds. These nearby, low-density clouds may not be in virial equilibrium and tend to deviate from the typical surface density observed in more evolved molecular clouds.

\begin{figure}
    \centering
    \includegraphics[width=1\linewidth]{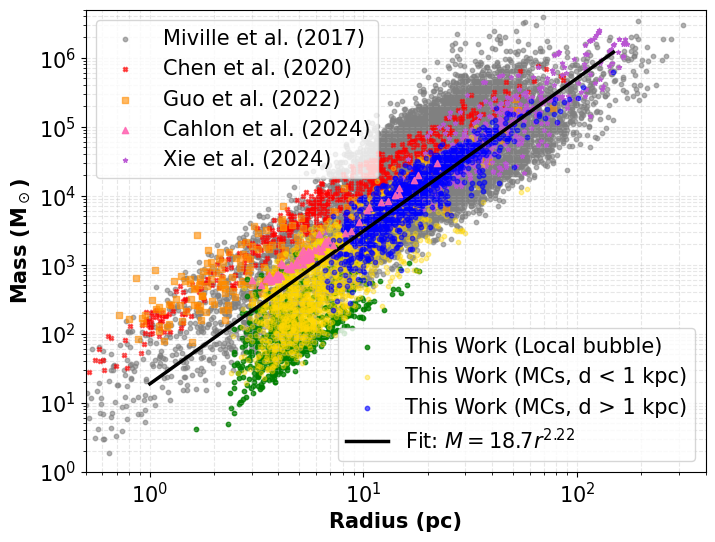}
    \caption{
    Mass-radius relation for molecular clouds in our catalog. Green circles: Local Bubble clouds; gold circles: non–Local Bubble clouds within 1 kpc; blue circles: clouds beyond 1 kpc. The black solid line shows the best-fit for the blue population. Data from previous studies are shown for comparison: \citet{Miville2017} (gray circles), \citet{chen2020} (red X markers), \citet{guo2022} (orange squares), \citet{Cahlon2024} (pink triangles), and \citet{xie2024} (purple stars).
    }
    \label{fig:relation}
\end{figure}

Figure~\ref{fig:relation} shows the mass-radius relation for our molecular cloud sample, with comparison to scaling laws from previous studies. In addition to the previously discussed studies, we now incorporate the mass and radius data extracted from Table 2 of \citet{Cahlon2024}. Their sample contains only 65 molecular clouds, and the data points are tightly clustered in parameter space, forming a relatively compact population. Due to the small sample size and narrow distribution in both mass and radius, we did not include this dataset in the histogram comparisons shown earlier. Nevertheless, the Cahlon et al. sample follows a remarkably tight linear trend, which makes it a useful point of reference for the mass–radius relation and thus is overplotted in Figure~\ref{fig:relation}. We find that clouds beyond 1\,kpc follow a power-law relation consistent with earlier results. Clouds within 1\,kpc also follow this trend but exhibit greater scatter, especially those associated with the Local Bubble. This increased dispersion is likely due to the inclusion of many nearby, low-mass clouds, which may not yet be gravitationally bound or in equilibrium.

\subsection{The Galactic Structure as Traced by Molecular Clouds}\label{sec:structure}

\begin{figure*}
    \centering
    \includegraphics[width=1\linewidth]{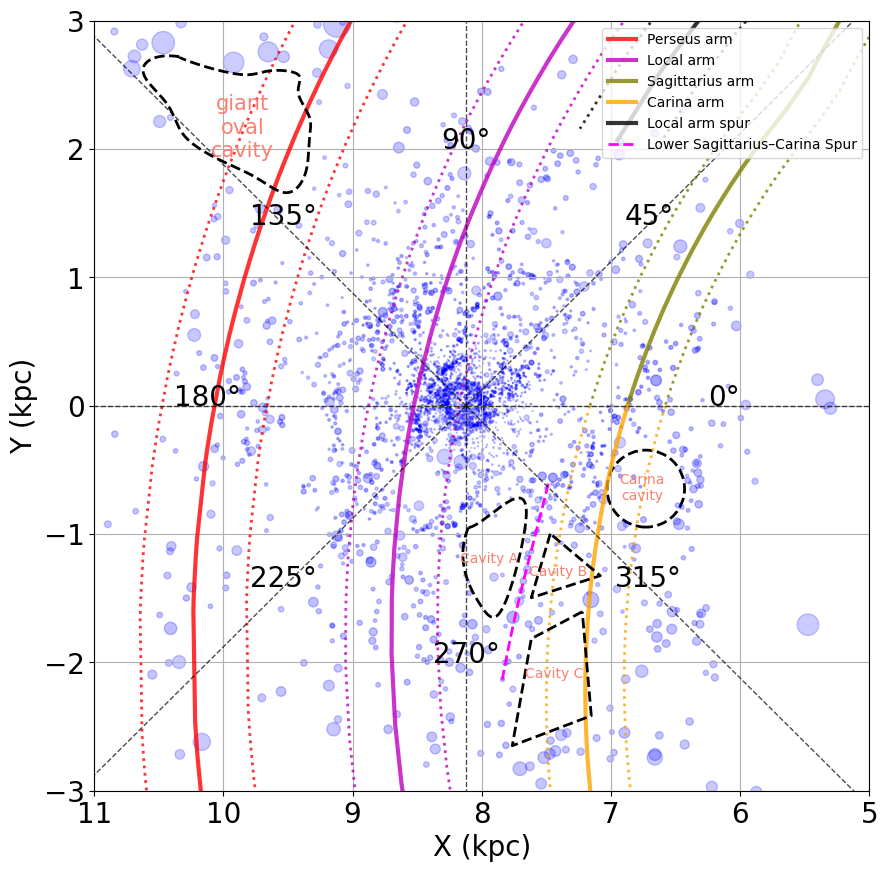}
    \caption{
    Spatial distribution of cataloged molecular clouds in the Galactic $X$-$Y$ plane. Each circle represents a molecular cloud, with radius proportional to its physical size and transparency indicating average dust density (a proxy for compactness). The Sun is at $(X, Y) = (8.12, 0)$\,kpc. Galactic longitudes are labeled for orientation. Solid and dashed curves represent spiral arm loci and their $\pm 1\sigma$ widths based on \citet{2019ApJ...885..131R}. Several large-scale cavities are outlined with black dashed lines, including the ``Giant Oval Cavity'' described by \citet{Vergely2022} and the ``Carina cavity'' described by \citet{chen2019a}. The ``Lower Sagittarius-Carina Spur'' described by \citet{chen2020} is outlined in purple.
    }
    \label{fig:MCXY}
\end{figure*}

Figure~\ref{fig:MCXY} presents the distribution of cataloged molecular clouds projected onto the Galactic $X$-$Y$ plane. The central Local Bubble cavity is clearly visible, characterized by a low density of molecular clouds surrounding the solar position.

In the second Galactic quadrant, we identify a prominent elliptical void that closely matches the ``Giant Oval Cavity'' reported by \citet{Vergely2022} (see their Figure 11). This structure is well defined in the molecular cloud distribution and supports the presence of an extended, low-density region in this part of the Galaxy.

Along the Sagittarius-Carina Arm, our analysis reveals a nearly circular region lacking molecular clouds, corresponding to the ``Carina cavity'' proposed by \citet{chen2019a} based on the spatial distribution of OB stars. The absence of clouds in this area provides additional evidence for the existence of a large-scale cavity within the arm.

The ``Lower Sagittarius-Carina Spur'', identified by \citet{chen2020}, appears more distinct and continuous in our data. This feature seems to form a bridge between the Local Arm and the Sagittarius-Carina Arm, suggesting a possible structural connection. The spur lies between two adjacent cavities, labeled ``Cavity A'' and ``Cavity B''. Additionally, ``Cavity B'' appears to connect with a deeper structure below, referred to as ``Cavity C'', forming a larger, coherent void.

\section{Summary} \label{sec:summary}

In this study, we present a new, comprehensive all-sky catalog of 3345 molecular clouds, identified from the W25 3D dust reddening map using a dendrogram-based clustering technique with distance-adaptive parameters. The catalog spans heliocentric distances from approximately 90\,pc to 4300\,pc and covers the full sky.

For each cloud, we provide measurements of position, distance, projected area, equivalent physical radius, total mass, surface mass density, and average dust density. About 650 clouds are associated with the boundary of the Local Bubble. Even after removing these, approximately 740 clouds remain at high Galactic latitudes ($|b| > 20^\circ$), most of which are nearby and faint.

The spatial distribution of the cataloged clouds reveals clear structural features of the interstellar medium. On local scales, the clouds trace the shell of the Local Bubble, highlighting its cavity-like morphology. On larger scales, we identify coherent spur-like structures and prominent large-scale cavities, which may be associated with the Galactic spiral arm network.

Our cloud sample provides a well-characterized dataset for studying the distribution of dust and gas in the solar neighborhood, the structure and dynamics of the Local Bubble, and the connection between molecular clouds and the large-scale morphology of the Milky Way. More detailed investigations of these topics will be pursued in future work.

\begin{acknowledgments}
The authors thank the anonymous referee for his/her suggestions that improved the quality and clarity of our presentation. This work is supported by the National Natural Science Foundation of China through the projects NSFC 12222301, 12173007, and 12173034, as well as the National Key Basic R\&D Program of China via 2024YFA1611901 and 2024YFA1611601.  

This work has made use of data products from the LAMOST and $\gaia$. 
Guoshoujing Telescope (the Large Sky Area Multi-Object Fiber Spectroscopic Telescope; LAMOST) is a National Major Scientific Project built by the Chinese Academy of Sciences. 
Funding for the project has been provided by the National Development and Reform Commission. 
LAMOST is operated and managed by the National Astronomical Observatories, Chinese Academy of Sciences. 

\end{acknowledgments}

\clearpage
\bibliography{ref}{}
\bibliographystyle{aasjournalv7}

\renewcommand{\thefigure}{S\arabic{figure}}
\setcounter{figure}{0}
\section*{Supplementary Material}
\label{supp}
\begin{figure*}
    \centering
    \includegraphics[width=1\linewidth]{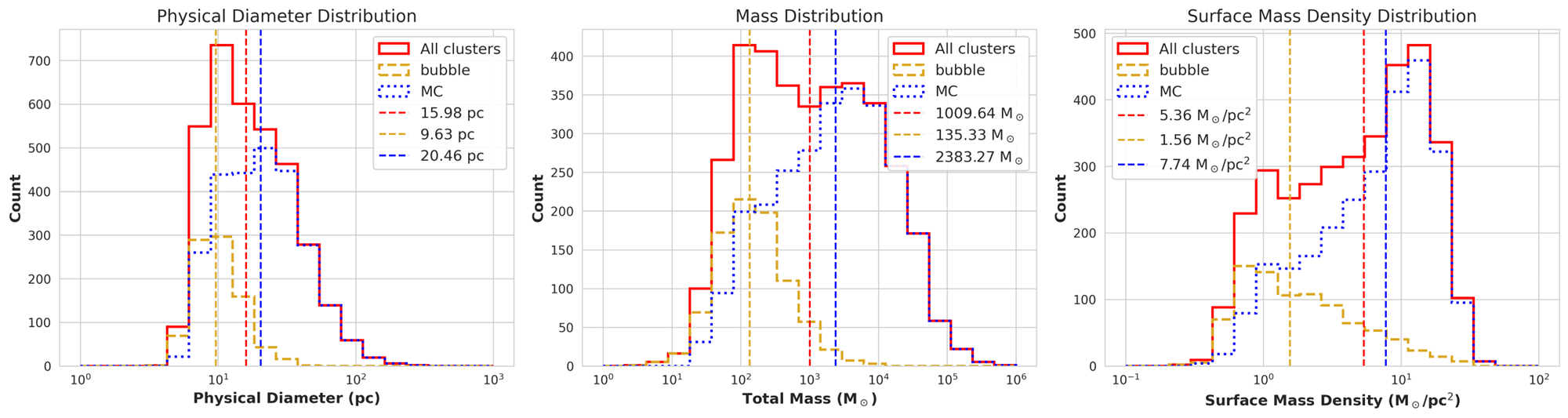}
    \includegraphics[width=1\linewidth]{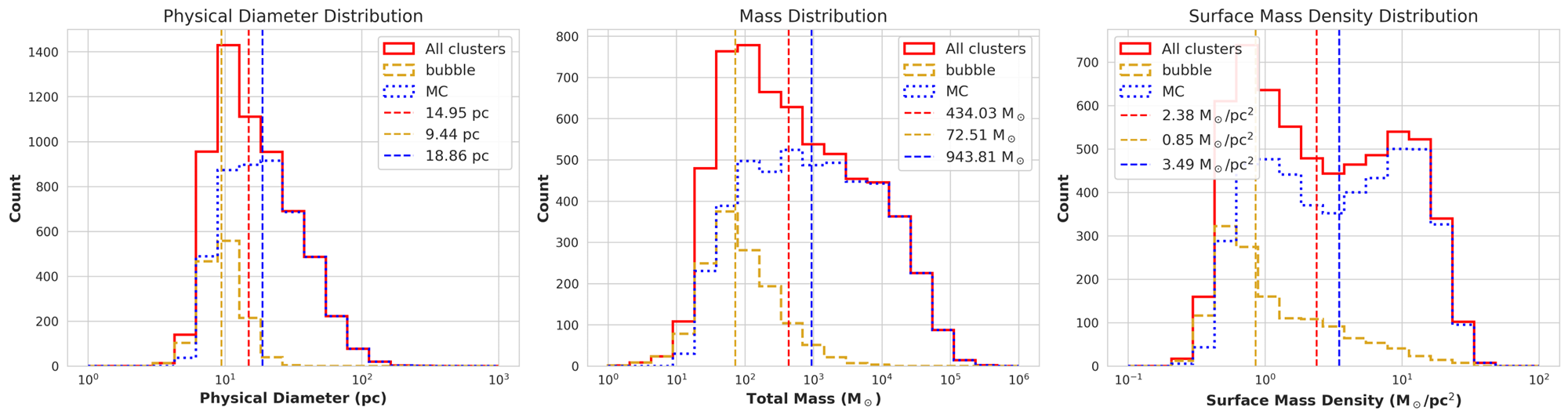}
    \includegraphics[width=1\linewidth]{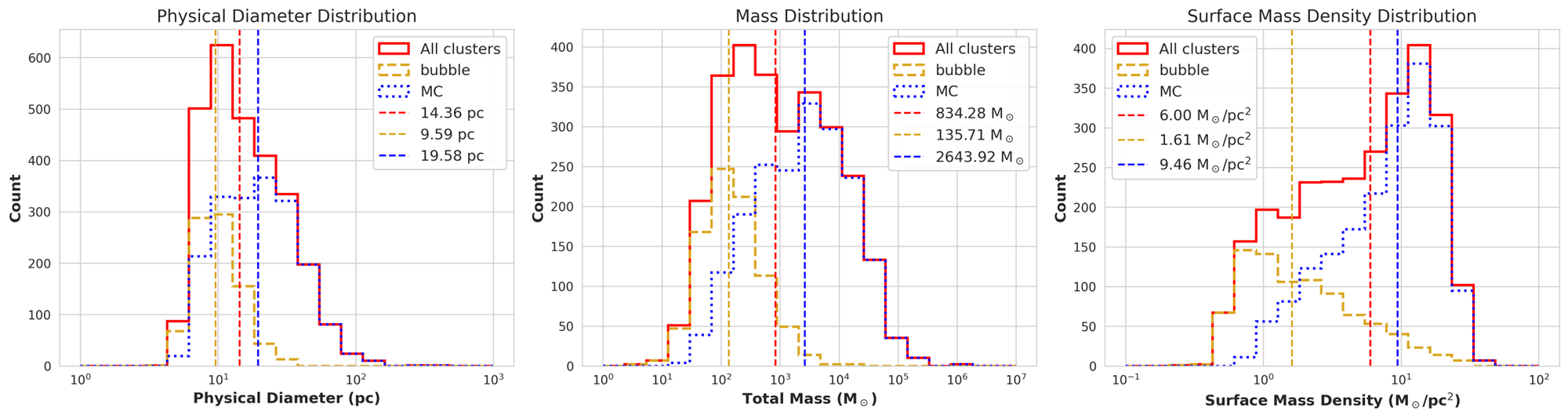}
    \includegraphics[width=1\linewidth]{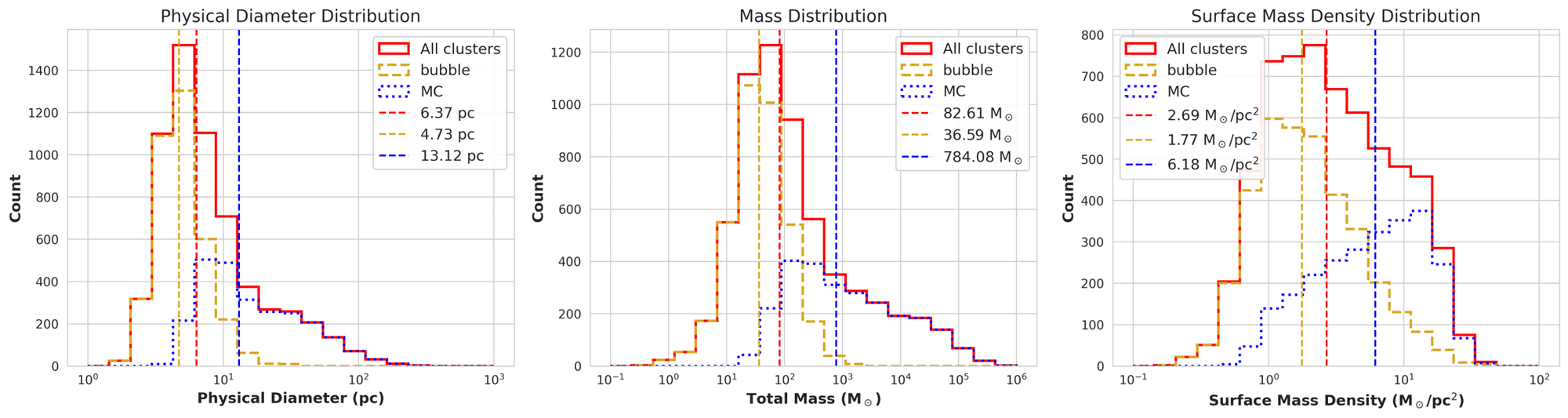}
    \caption{Histograms of cloud physical properties under different clustering parameter configurations. From top to bottom, the four panels show results obtained by modifying a single parameter from the baseline values listed in Table~\ref{tab:param}: (1) setting all \texttt{min\_value} values to 0.01~mag/kpc; (2) setting all \texttt{min\_delta} values to 0.1~mag/kpc; (3) setting all \texttt{min\_delta} values to 0.5~mag/kpc; and (4) setting all \texttt{min\_npix} values to 500.}
    \label{fig:param1234}
\end{figure*}

\begin{figure*}
    \centering
    \includegraphics[width=1\linewidth]{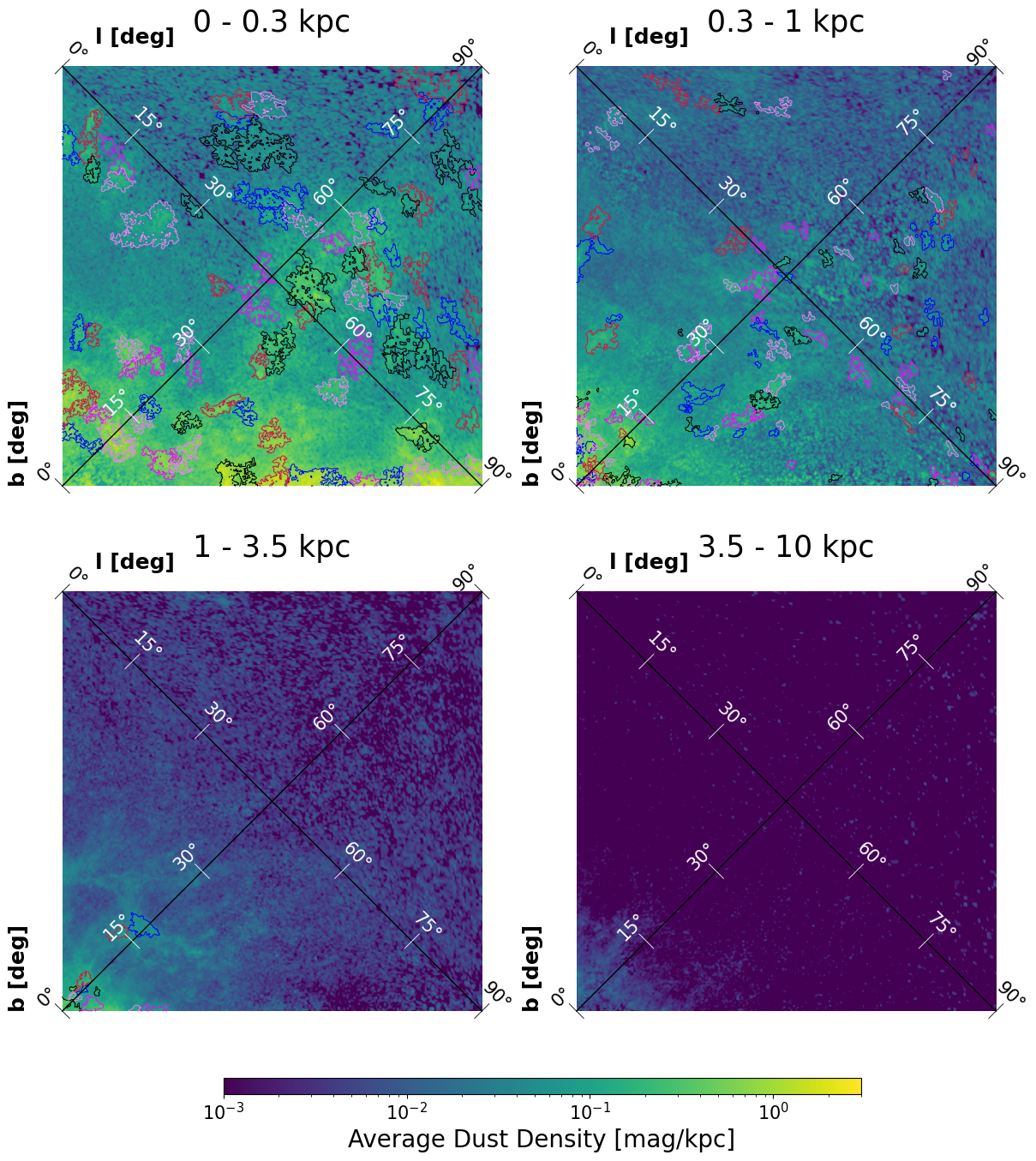}
    \caption{This figure presents the clustering results for the region corresponding to HEALPix pixel index 0 in Figure \ref{fig:HEALPix}, spanning Galactic longitudes from $0^\circ$ to $90^\circ$ and latitudes $0^\circ$ to $90^\circ$. The four subpanels correspond to different distance intervals: 0--0.3\,kpc, 0.3--1\,kpc, 1--3.5\,kpc, and 3.5--10\,kpc. Molecular clouds identified in each bin are outlined accordingly. The \texttt{Dendrograms} parameters used in each distance interval are listed in Table~\ref{tab:param}. Distinct clouds along the line of sight are represented using different colors.}
    \label{fig:pix0}
\end{figure*}

Figure~\ref{fig:param1234} shows histograms of the physical properties of identified clouds under several different sets of clustering parameters. This supplementary material also presents clustering results for the ten additional sky regions not shown in the main text, completing the full set of twelve regions into which the W25 extinction map was divided. All regions were processed using the same \texttt{Dendrograms}-based approach described in Section~\ref{sec:method}.

\begin{figure*}
    \centering
    \includegraphics[width=1\linewidth]{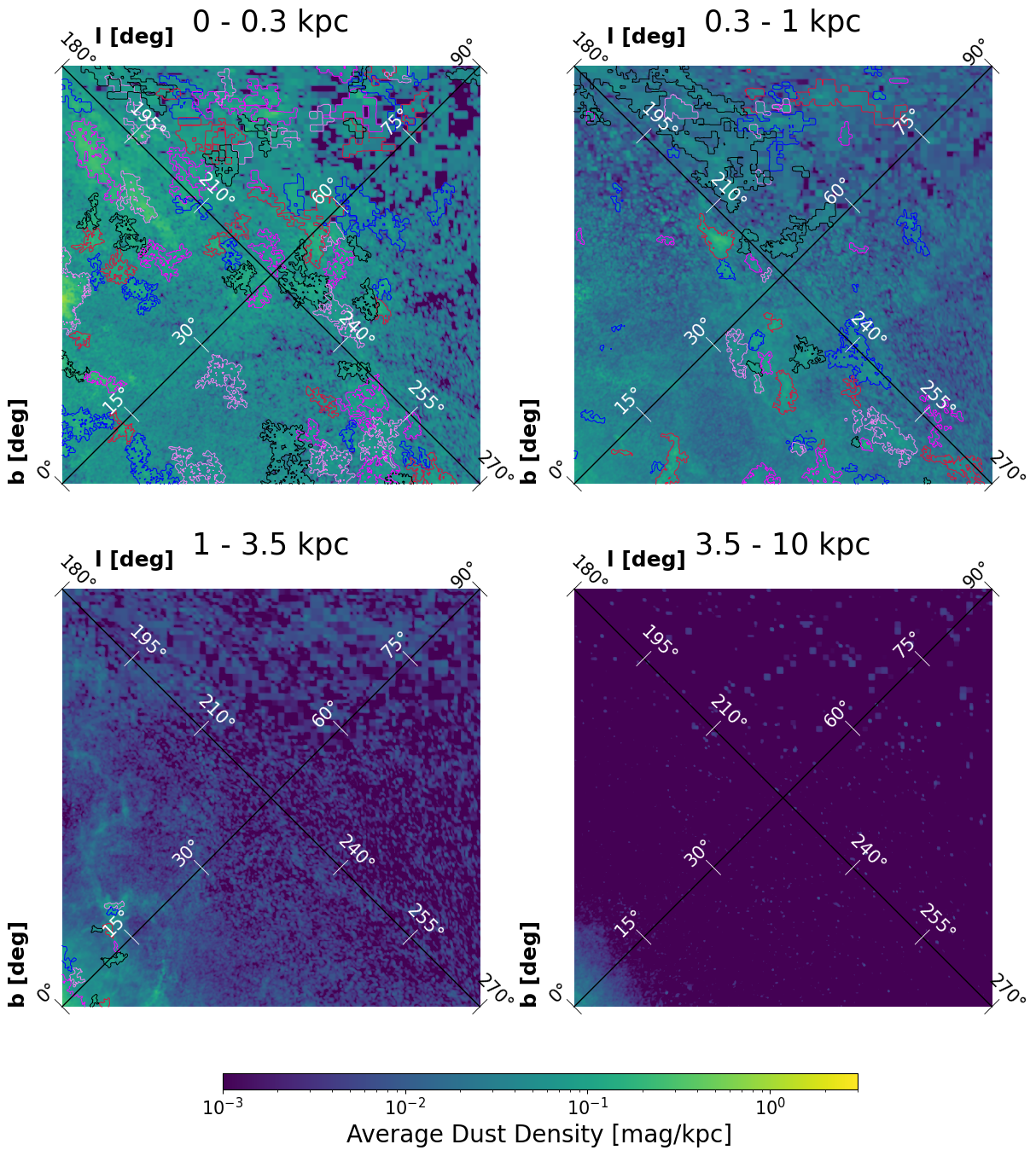}
    \caption{Similar to Figure\,\ref{fig:pix0} but for the regio spanning Galactic longitudes $180^\circ$ to $270^\circ$ and latitudes $0^\circ$ to $90^\circ$, corresponding to HEALPix pixel index 2 in Figure~\ref{fig:Results-example1}. The low-resolution regions near the top of each subpanel arise from the adaptive spatial binning in W25, which is based on local stellar density. These areas correspond to sky regions with relatively sparse $Gaia$ coverage, resulting in reduced angular resolution.
    }
    \label{fig:pix2}
\end{figure*}
\begin{figure*}
    \centering
    \includegraphics[width=1\linewidth]{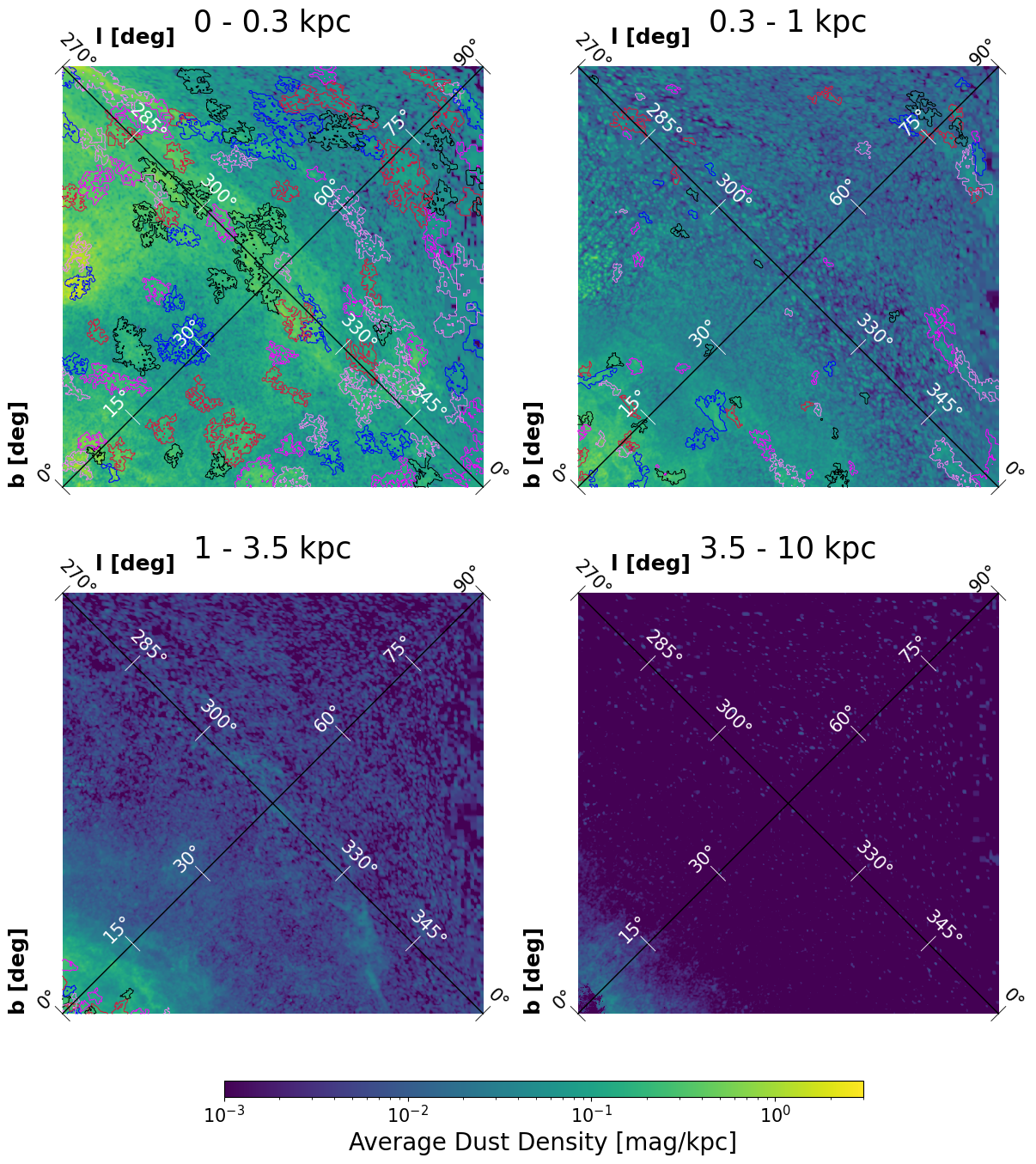}
    \caption{Similar to Figure\,\ref{fig:pix0} but for the region spanning Galactic longitudes $270^\circ$ to $360^\circ$ and latitudes $0^\circ$ to $90^\circ$, corresponding to HEALPix pixel index 3 in Figure~\ref{fig:Results-example1}.}
    \label{fig:pix3}
\end{figure*}
\begin{figure*}
    \centering
    \includegraphics[width=1\linewidth]{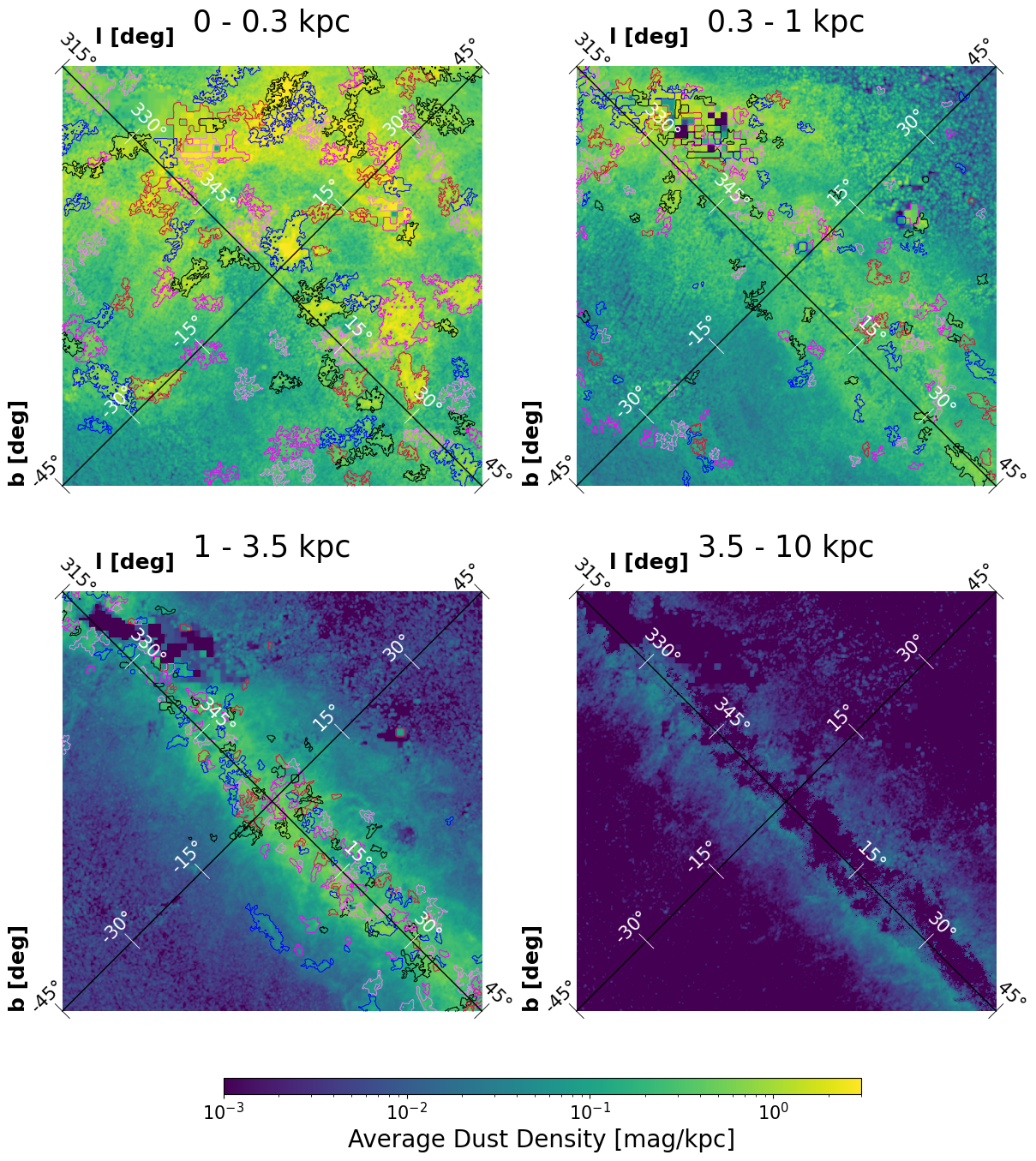}
    \caption{Similar to Figure\,\ref{fig:pix0} but for the region spanning Galactic longitudes $-45^\circ$ to $45^\circ$ and latitudes $-45^\circ$ to $45^\circ$, corresponding to HEALPix pixel index 4 in Figure~\ref{fig:Results-example1}. The low-resolution region near the upper center of each subpanel, corresponding to the Galactic center direction, arises from the adaptive angular binning scheme used in W25, which is based on local stellar density. In this area, heavy dust extinction reduces Gaia’s detection depth, resulting in lower stellar densities and consequently coarser angular resolution.}
    \label{fig:pix4}
\end{figure*}
\begin{figure*}
    \centering
    \includegraphics[width=1\linewidth]{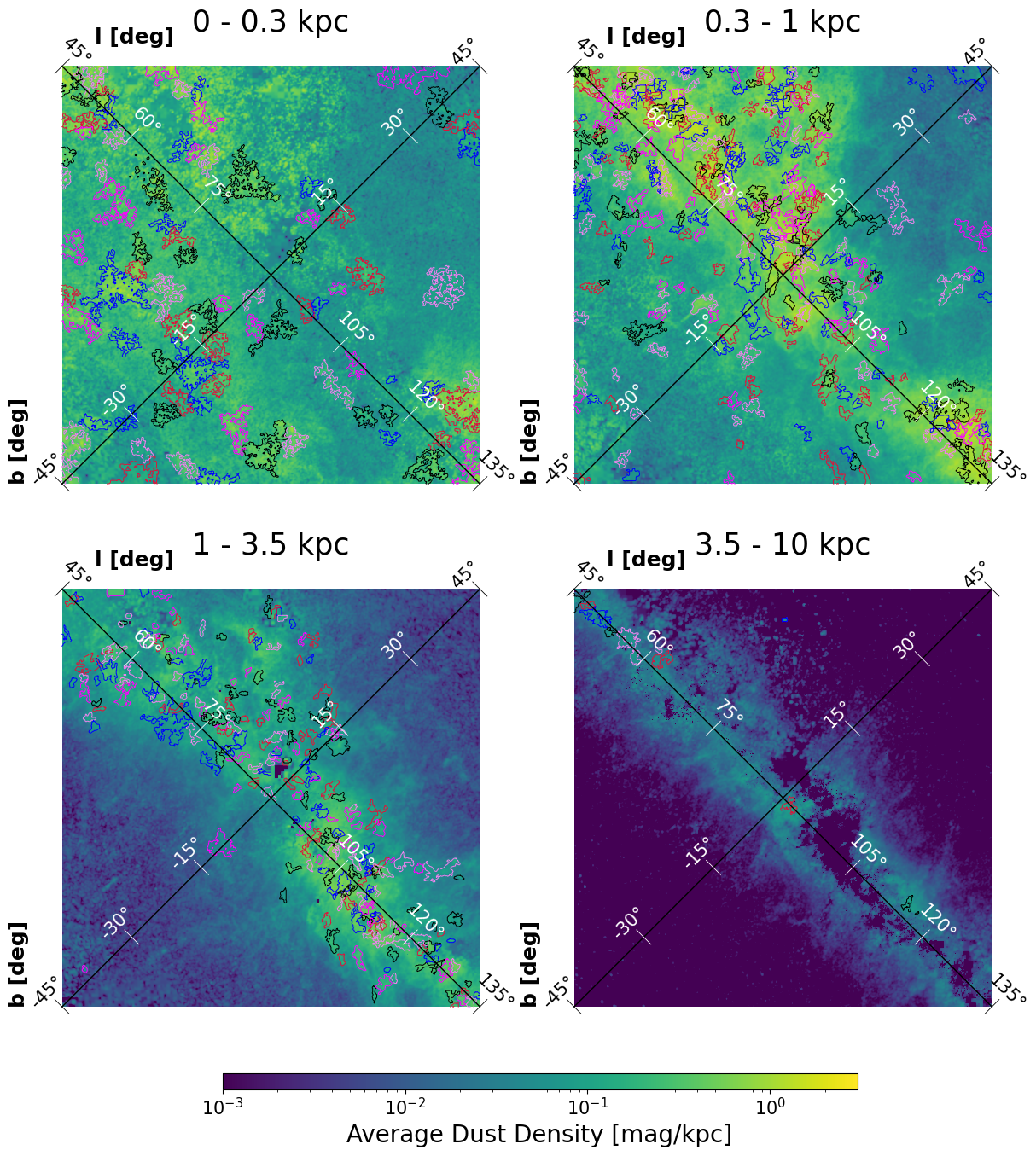}
    \caption{Similar to Figure\,\ref{fig:pix0} but for the region spanning Galactic longitudes $45^\circ$ to $135^\circ$ and latitudes $-45^\circ$ to $45^\circ$, corresponding to HEALPix pixel index 5 in Figure~\ref{fig:Results-example1}.}
    \label{fig:pix5}
\end{figure*}
\begin{figure*}
    \centering
    \includegraphics[width=1\linewidth]{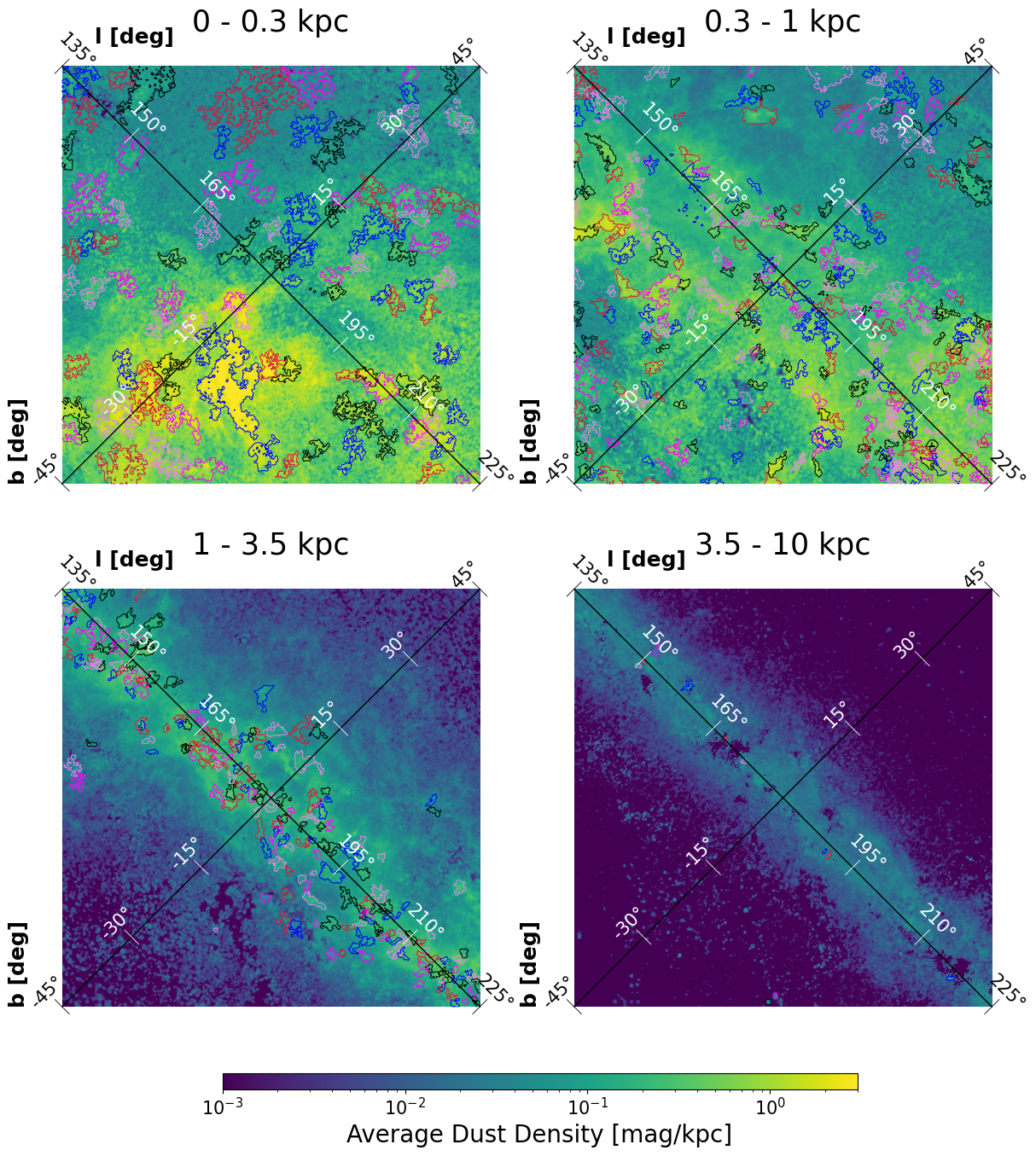}
    \caption{Similar to Figure\,\ref{fig:pix0} but for the region spanning Galactic longitudes $135^\circ$ to $225^\circ$ and latitudes $-45^\circ$ to $45^\circ$, corresponding to HEALPix pixel index 6 in Figure~\ref{fig:Results-example1}.}
    \label{fig:pix6}
\end{figure*}
\begin{figure*}
    \centering
    \includegraphics[width=1\linewidth]{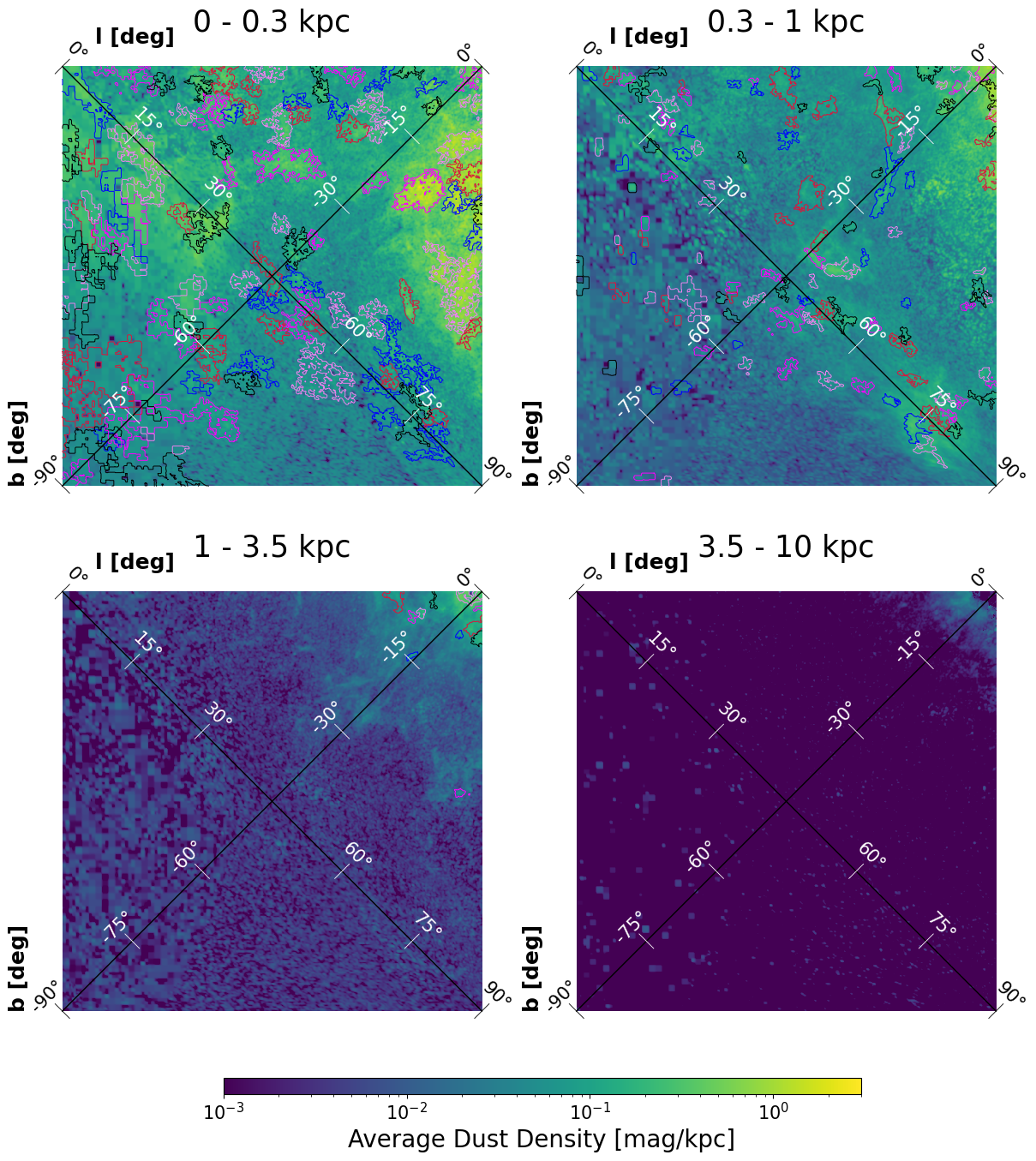}
    \caption{Similar to Figure\,\ref{fig:pix0} but for the region spanning Galactic longitudes $0^\circ$ to $90^\circ$ and latitudes $-90^\circ$ to $0^\circ$, corresponding to HEALPix pixel index 8 in Figure~\ref{fig:Results-example1}. The low-resolution regions near the bottom of each subpanel arise from the adaptive spatial binning in W25, which is based on local stellar density. These areas correspond to sky regions with relatively sparse $Gaia$ coverage, resulting in reduced angular resolution.}
    \label{fig:pix8}
\end{figure*}
\begin{figure*}
    \centering
    \includegraphics[width=1\linewidth]{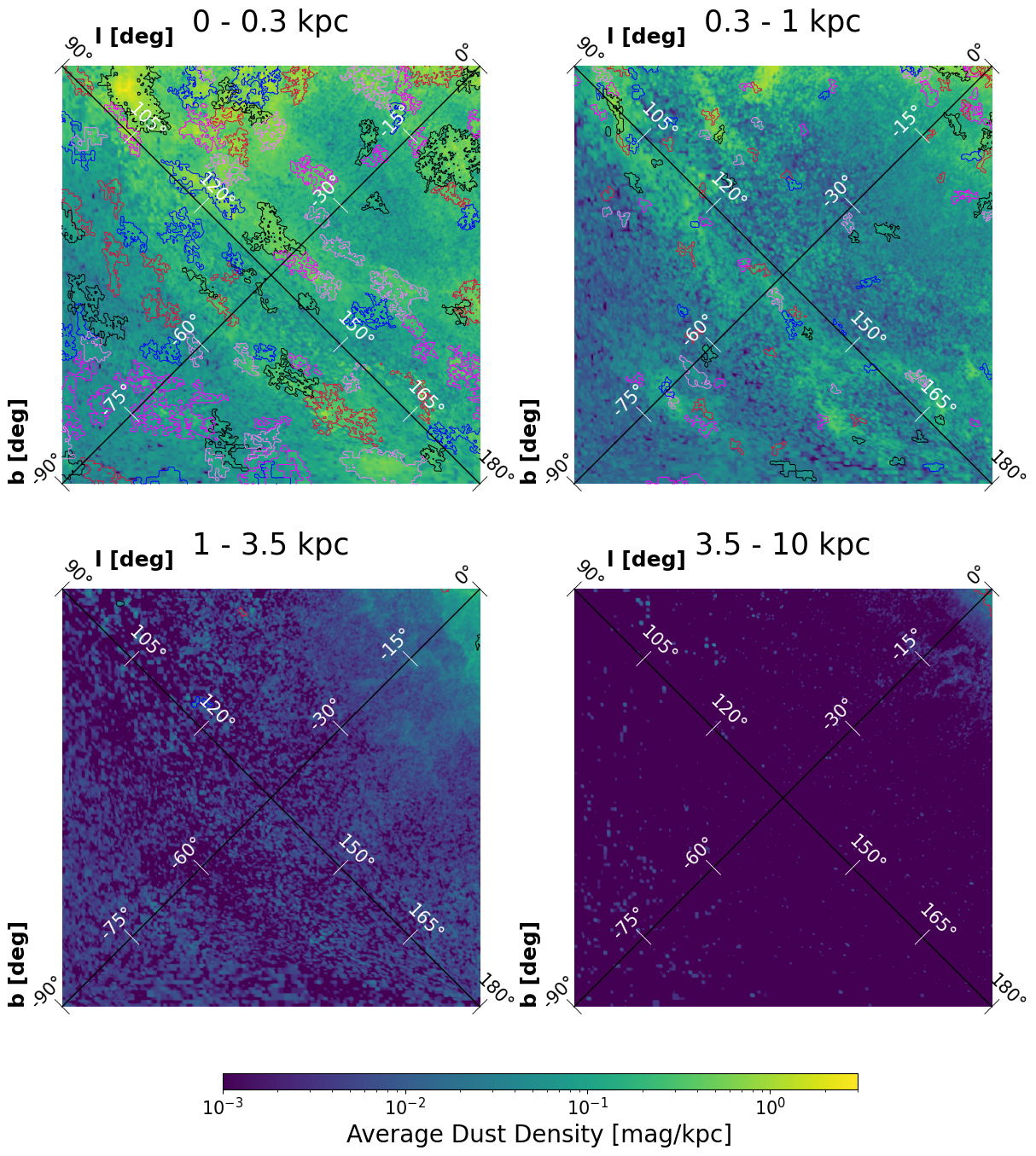}
    \caption{Similar to Figure\,\ref{fig:pix0} but for the region spanning Galactic longitudes $90^\circ$ to $180^\circ$ and latitudes $-90^\circ$ to $0^\circ$, corresponding to HEALPix pixel index 9 in Figure~\ref{fig:Results-example1}.}
    \label{fig:pix9}
\end{figure*}
\begin{figure*}
    \centering
    \includegraphics[width=1\linewidth]{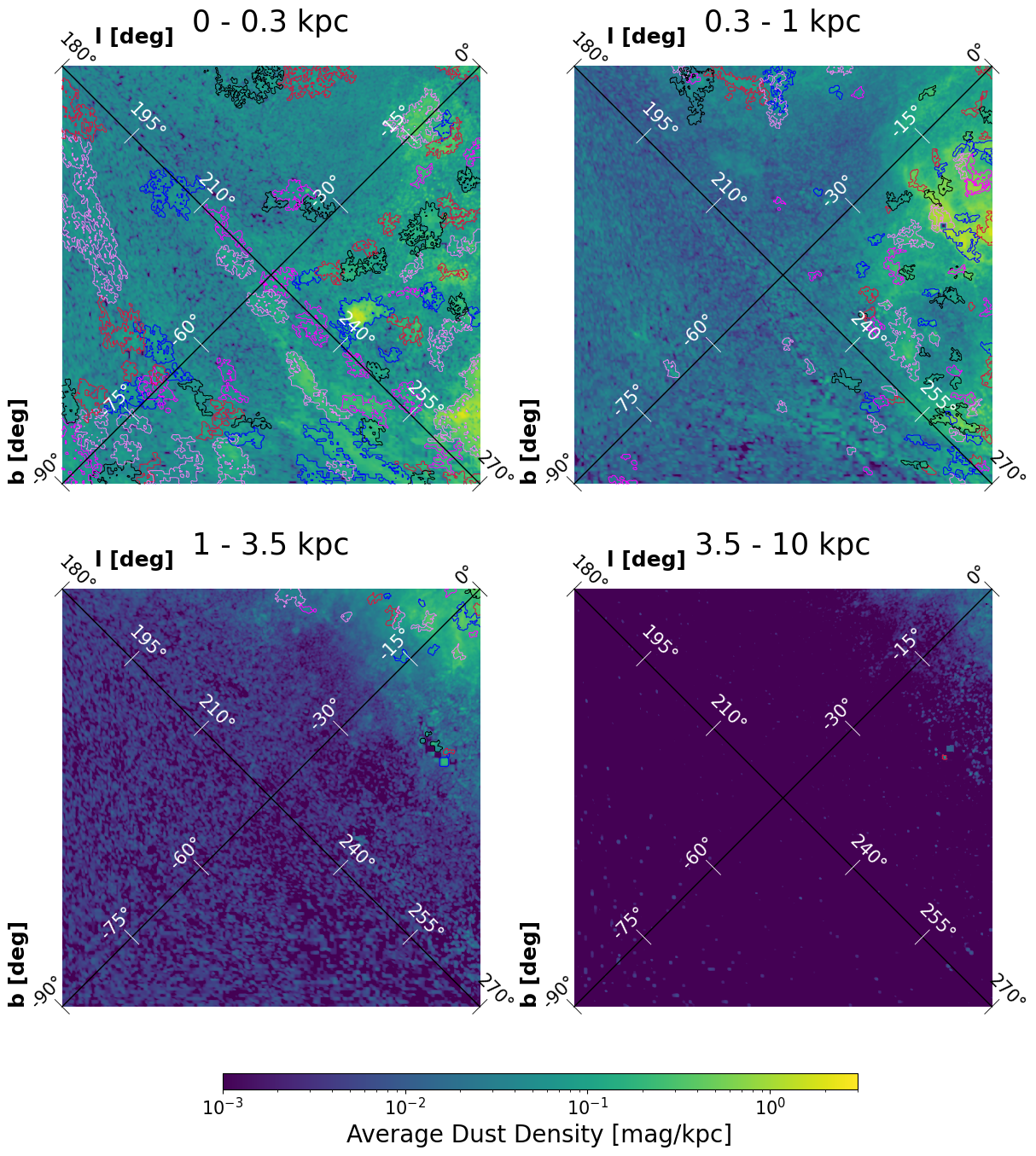}
    \caption{Similar to Figure\,\ref{fig:pix0} but for the region spanning Galactic longitudes $180^\circ$ to $270^\circ$ and latitudes $-90^\circ$ to $0^\circ$, corresponding to HEALPix pixel index 10 in Figure~\ref{fig:Results-example1}.}
    \label{fig:pix10}
\end{figure*}
\begin{figure*}
    \centering
    \includegraphics[width=1\linewidth]{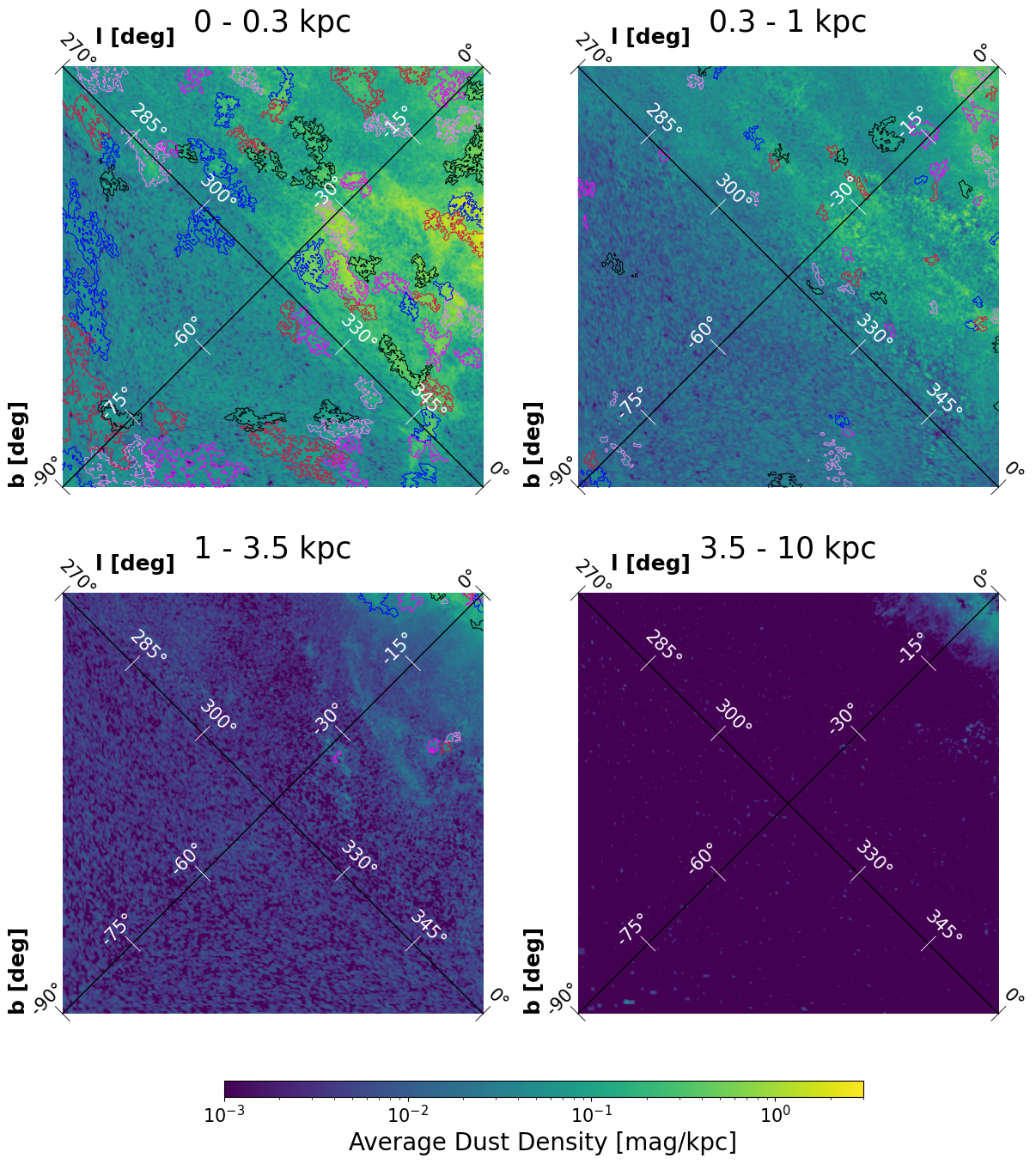}
    \caption{Similar to Figure\,\ref{fig:pix0} but for the region spanning Galactic longitudes $270^\circ$ to $360^\circ$ and latitudes $-90^\circ$ to $0^\circ$, corresponding to HEALPix pixel index 11 in Figure~\ref{fig:Results-example1}.}
    \label{fig:pix11}
\end{figure*}

\end{CJK*}
\end{document}